\pdfoutput=1 

\documentclass[journal]{IEEEtran}
\usepackage{lineno,hyperref}
\usepackage{amsmath}
\usepackage{indentfirst}
\usepackage{booktabs}
\usepackage{subfigure}
\usepackage{epsfig}
\usepackage{amssymb}
\usepackage{graphicx}
\usepackage{float}
\usepackage{setspace}
\usepackage{hyperref} 
\hypersetup{colorlinks=true,urlcolor=blue,citecolor=blue,linkcolor=blue,bookmarks=true} 
\usepackage{tabularx,booktabs}
\usepackage{url}
\usepackage{graphicx}
\usepackage[table,xcdraw]{xcolor}
\usepackage{mathrsfs}
\usepackage{cite}%
\usepackage[center]{caption} %
\usepackage{enumerate}%
\input{Qcircuit.tex} %

\usepackage{algorithmic}
\usepackage{algorithm}

\ifCLASSINFOpdf
\else
\fi
\begin{document}
%
\title{QZNs: Quantum Z-numbers}
%
%
%

\author{Jixiang Deng,
	Yong Deng*
	\thanks{Jixiang Deng is with Institute of Fundamental and Frontier Science, University of Electronic Science and Technology of China, Chengdu, 610054, China.}
	\thanks{Yong Deng is with Institute of Fundamental and Frontier Science, University of Electronic Science and Technology of China, Chengdu, 610054, China; School of Education, Shannxi Normal University, Xi'an, 710062, China; School of Knowledge Science, Japan Advanced Institute of Science and Technology, Nomi, Ishikawa 923-1211, Japan.}
	\thanks{Corresponding author: Yong Deng. 
	 (e-mail: dengentropy@uestc.edu.cn;  prof.deng@hotmail.com)}
}

%
%

\markboth{Journal of \LaTeX\ Class Files,~Vol.~14, No.~8, August~2015}%
{Shell \MakeLowercase{\textit{et al.}}: Bare Demo of IEEEtran.cls for IEEE Journals}
%



\maketitle

\begin{abstract}
Because of the efficiency of modeling fuzziness and vagueness, Z-number plays an important role in real practice. However, Z-numbers, defined in the real number field, lack the ability to process the quantum information in quantum environment. It is reasonable to generalize Z-number into its quantum counterpart. In this paper, we propose quantum Z-numbers (QZNs), which are the quantum generalization of Z-numbers. In addition, seven basic quantum fuzzy operations of QZNs and their corresponding quantum circuits are presented and illustrated by numerical examples. Moreover, based on QZNs, a novel quantum multi-attributes decision making (MADM) algorithm is proposed and applied in medical diagnosis. The results show that, with the help of quantum computation, the proposed algorithm can make diagnoses correctly and efficiently.
\end{abstract}

\begin{IEEEkeywords}
Z-numbers, quantum computation,  quantum Z-numbers, quantum fuzzy operations, quantum multi-attributes decision making algorithm.
\end{IEEEkeywords}%

%
\IEEEpeerreviewmaketitle

\section{Introduction}
\IEEEPARstart{I}{n} the last couple of decades, for modeling and handling the uncertain information in the different environment, many theories have been developed, such as probability theory \cite{lee1980probability}, fuzzy set theory \cite{zadeh1965fuzzy1, klir1995fuzzy}, evidence theory \cite{dempster1967,Shafer1976}, complex evidence theory  \cite{Xiao2019complexmassfunction,Xiao2020CEQD},   belief structure  \cite{Yager2017MaxitiveBelief,Yager2018Fuzzygeneralizedbeliefstructure},  Z-numbers \cite{zadeh2011note}, and D numbers \cite{LiuPD2020132,LiubyDFMEA}.

%

For representing and processing vagueness  in the real world, fuzzy set theory, proposed by Zadeh in 1965 \cite{zadeh1965fuzzy1}, attracts much attentions and is widely used in many field \cite{dzitac2017fuzzy}. In order to better dealing with information in fuzzy environment, fuzzy set theory is extensively developed, and modifications of fuzzy sets continuously emerge, e.g., type-2 fuzzy sets \cite{zadeh1975concept}, intuitionistic fuzzy sets  \cite{atanassov1999intuitionistic}, and Pythagorean fuzzy sets \cite{yager2013pythagorean1}. However, classical fuzzy set just models the fuzziness of uncertain information, but does not take the   reliability of the processed information into consideration. Hence, considering both fuzziness and reliability, Zadeh proposed Z-numbers \cite{zadeh2011note}. Lots of researches further promote the development of Z-numbers, such as  arithmetic of Z-numbers \cite{aliev2015arithmetic,aliev2016arithmetic}, applied model of Z-numbers \cite{patel2015applied}, ranking  of Z-numbers \cite{aliev2016ranking,bakar2015multi},   and  uncertainty measurements of Z-numbers \cite{kang2018method,wang2017multi}. Based on these methods, Z-number has been broadly applied in various areas, including  approximate reasoning \cite{aliev2016approximate}, expert system  \cite{tian2020zslf}, data fusion \cite{jiang2016sensor},   medical diagnosis \cite{wu2018new},  environmental assessment \cite{kang2020environmental}, and decision making \cite{kang2012decision,aliev2014decision,wang2017multi}.

Blessed with the properties of quantum mechanics such as  quantum entanglement and quantum parallelism, quantum algorithms have long been known for the speedups of several problems such as factoring \cite{shor1994algorithms} and searching \cite{grover1996fast}. Recently, quantum mechanics and quantum computation  draw more and more attentions \cite{nielsen2002quantum,schuld2015introduction}. Many quantum algorithms emerge and lots of classical algorithms have been generalized into their quantum counterparts, such as quantum neural networks \cite{kak1995quantum,cong2019quantum}, quantum support vector machine \cite{rebentrost2014quantum}, quantum reinforcement learning \cite{dong2008quantum}, quantum Parrondo's games \cite{lai2020parrondoeffect,lai2020parrondoparadox,lai2020parrondowalk},  and so on \cite{wiebe2014quantum, gao2020quantum, gao2021quantum}.

As for the quantum generalization of fuzzy set theory,  Pykacz proposed quantum fuzzy logic \cite{pykacz1992fuzzy, pykacz1994fuzzy}  and Mannucci presented quantum fuzzy sets \cite{mannucci2006quantum}. In 2016, Reiser extended  intuitionistic fuzzy set into its quantum model \cite{reiser2016interpretations}. However, Z-numbers  are defined in the real number field \cite{zadeh2011note}, which lack the ability of representing quantum information, so that the existing algorithm based on Z-numbers cannot be applied in  quantum computation. Hence, it is reasonable to extend Z-numbers into   Hilbert space and propose the quantum model of Z-numbers.

In this paper,  we propose quantum Z-numbers (QZNs), which are the quantum generalization of classical Z-numbers. A QZN consists of two quantum fuzzy sets. The first set is a quantum fuzzy restriction, and the second set measures the reliability of the first set.  In addition, several basic quantum fuzzy  operations  of QZNs and their associated quantum circuits are proposed. Numerical examples are shown to illustrate QZN and its operations. Moreover, we present a novel quantum multi-attributes decision making (MADM) algorithm based on QZNs, which is applied in medical diagnosis. The results show that, duo to the advantage of quantum computation,  the proposed algorithm can efficiently dealing with fuzziness and correctly make diagnoses with low time complexity. 

The major contributions of this paper are as follows.
\begin{enumerate}[(1)]
	\item Quantum Z-numbers (QZNs) are proposed, which are the quantum extension of classical Z-numbers.
	\item Seven quantum fuzzy operations of QZNs are proposed, which are illustrated by numerical examples.
	\item A new quantum MADM algorithm is designed based on QZNs, and its advantage of time complexity in big data scenario is analyzed. 
	\item The proposed algorithm is applied in medical diagnosis, which shows its efficiency of handling fuzziness and the correctness of make diagnoses with low time complexity.
\end{enumerate}

The rest of this paper is organized as follows. Section  \uppercase\expandafter{\romannumeral2} briefly reviews some preliminaries. Section  \uppercase\expandafter{\romannumeral3} proposes the  quantum Z-numbers (QZNs) and the quantum fuzzy  operations of QZNs. Some numerical examples are illustrated in Section \uppercase\expandafter{\romannumeral4}. Section  \uppercase\expandafter{\romannumeral5} proposes a quantum MADM algorithm based on QZNs and analyzes the time complexity of the proposed algorithm.   In Section \uppercase\expandafter{\romannumeral6}, the proposed algorithm is applied in medical diagnosis, and the results are analyzed and discussed. Section \uppercase\expandafter{\romannumeral7} makes a brief conclusion.

\section{Preliminaries}

Several preliminaries are briefly introduced in this section, including fuzzy sets, Z-numbers, three operations of fuzzy sets, quantum mechanics and quantum computation, quantum fuzzy sets and quantum fidelity.  

\subsection{Fuzzy sets}

Fuzzy set  \cite{zadeh1965fuzzy1} is an efficient tool for processing fuzziness in different environment. Researchers have presented a lot of methods based on fuzzy sets, such as fuzzy distance \cite{xiao2019distance}, fuzzy divergence \cite{xiao2019divergence,zhou2020new}, fuzzy similarity \cite{xu2009intuitionistic}, fuzzy entropy \cite{NguyenXuan2019}, and fuzzy information volume \cite{deng2021Informationvolume}.

Let $U$ be a universe of discourse. A fuzzy set $A$ based on $U$ is a set of pairs  defined as \cite{zadeh1965fuzzy1}:

\noindent\textbf{Definition 2.1:}
\emph{ Fuzzy sets}

\emph{
	\begin{equation}
	A=\{
	\left< x,\mu\left( x \right) \right> | x\in U \}
	\end{equation}
	where $\mu(x) :U\rightarrow \left[ 0,1 \right] $ is the membership function (MF) of $A$, describing the membership degree of each element $x$ to the fuzzy set $A$. 
}

\subsection{Z-numbers}

Given a universe of discourse $U$, a Z-number is an ordered pair of fuzzy numbers defined as follows \cite{zadeh2011note}:

\noindent\textbf{Definition 2.2:}
\emph{ Z-numbers}

\emph{
	\begin{equation}
	Z=(A, B)
	\end{equation}
	where the first component $A$ is a fuzzy restriction on the real-valued uncertain variable $x \in U$, and the second component $B$ is a measure the reliability of component $A$.
}

\subsection{Typical operations of fuzzy sets}

In this subsection, several operation of fuzzy sets will be introduced, including fuzzy complements, fuzzy intersections, and fuzzy unions.

Given $x,y,z\in[0,1]$, fuzzy complements, fuzzy intersections, and fuzzy unions are defined as follows \cite{klir1995fuzzy}: 

\noindent\textbf{Definition 2.3:}
\emph{ Fuzzy complements}

\emph{A fuzzy complement is an operation $C: [0,1] \rightarrow [0,1]$ which satisfies: (1) $C(0)=1$ and $C(1)=0$; (2) if $x\le y$, then $C(x) \ge C(y)$.
}

\noindent\textbf{Definition 2.4:}
\emph{ Fuzzy intersections (t-norms)}

\emph{A fuzzy intersection (also called t-norm) is an operation $I: [0,1]^2 \rightarrow [0,1]$ which satisfies: (1) $I(x,1)=x$; (2) if $y\le z$, then $I(x,y) \le I(x,z)$; (3) $I(x,y)=I(y,x)$; (4)  $I(x,I(y,z))=I(I(x,y),z)$.
}

\noindent\textbf{Definition 2.5:}
\emph{ Fuzzy unions (t-conorms) }

\emph{A fuzzy union (also called t-conorm) is an operation $U: [0,1]^2 \rightarrow [0,1]$ which satisfies: (1) $U(x,0)=x$; (2) if $y\le z$, then $U(x,y) \le U(x,z)$; (3) $U(x,y)=U(y,x)$; (4)  $U(x,U(y,z))=U(U(x,y),z)$.
}

There are many different kinds of fuzzy complements, fuzzy intersections, and fuzzy unions. In this paper, the following operations are taken into consideration \cite{klir1995fuzzy}: classical complement $C(x)=1-x$, algebraic product $I(x,y)=xy$, and algebraic sum $U(x,y)=x+y-xy$, where $x,y\in [0,1]$.

\subsection{Quantum mechanics and quantum computation}

According to \cite{nielsen2002quantum}, quantum mechanics is a mathematical framework for the development of physical theories. There are four postulates of quantum mechanics. 

Firstly, complex vector space with inner product, namely Hilbert space denoted as $\mathcal{H}$, is associated to any isolated physical system, which is known as the state space of the system. Quantum mechanical system can be fully described by state vector in Hilbert space. A  qubit is the simplest quantum system, which can be described by superposition state
\begin{equation}
	\ket{\psi} = \alpha\ket{0}+\beta\ket{1} 
\end{equation}
where $\ket{0}= [1,0]^T $ and $\ket{1} = [0,1]^T $  ($T$ denotes the transpose) are the orthonormal basis. $\alpha$ and $\beta$ are complex numbers, which satisfy the normalization condition $\left< \psi|\psi  \right> =\left| \alpha \right|^2+\left| \beta  \right|^2=1$.

Secondly, the evolution of a closed quantum system from state $\ket{\psi}$ to $\ket{\psi'}$ is described by a unitary transformation 
\begin{equation}
\ket{\psi'} = U \ket{\psi}
\end{equation}
where $U$ is a unitary operator which satisfies $U^\dag U=I$. In quantum computation, every unitary operator has its associated  quantum logic gate, which can be represented by quantum circuit symbol and matrix representation.   
Several basic quantum gates  that this paper considers are summarize in  TABLE \ref{TABLE.1}. For better understanding, let the input be  $\ket{x_i}$ and the output be $\ket{y_i}$ shown in TABLE \ref{TABLE.1}. The truth  tables of CCNOT gate and CSWAP gate are given in TABLE  \ref{TABLE.2}.

\begin{table}[t]
	\caption{Basic quantum logic gates, circuit symbols, and their  matrix representations}
	\centering
	\begin{tabular}{c|c|c}
		\hline
		Quantum gate &  Circuit symbol  & Matrix representation\\
		\hline
		& & \\
		Hadamard gate 
		&
		$\Qcircuit @C=1em @R=.7em
		{
			\lstick{} & \gate{H}  & \rstick{} \qw 
		}$
		&
		$\frac{1}{\sqrt{2}}\left[ \begin{smallmatrix}
		1&		1\\
		1&		-1\\
		\end{smallmatrix} \right] $ \\
		& & \\
		Pauli-X gate &  $\Qcircuit @C=1em @R=.7em
		{
			\lstick{} & \gate{X}  & \rstick{} \qw 
		}$ &  $
		\left[ \begin{smallmatrix}
		0&		1\\
		1&		0\\
		\end{smallmatrix} \right] 
		$\\
		& & \\
		Y-Rotation gate &  $\Qcircuit @C=1em @R=.7em
		{
			\lstick{} & \gate{R_Y^\theta}  & \rstick{} \qw 
		}$ &  $\left[ \begin{smallmatrix}
		\cos \frac{\theta}{2}&		-\sin \frac{\theta}{2}\\
		\sin \frac{\theta}{2}&		\cos \frac{\theta}{2}\\
		\end{smallmatrix} \right] 
		$\\
		& & \\
		CCNOT gate &  $\Qcircuit @C=0.6em @R=0.6em
		{
			\lstick{\ket{x_1}} & \ctrl{1}  & \rstick{\ket{y_1}}                 \qw \\
			\lstick{\ket{x_2}} & \ctrl{1}  & \rstick{\ket{y_2}}                 \qw \\
			\lstick{\ket{x_3}} & \targ      & \rstick{\ket{y_3}} \qw
		}$ &  $\left[ \begin{smallmatrix}
		1&		0&		0&		0&		0&		0&		0&		0\\
		0&		1&		0&		0&		0&		0&		0&		0\\
		0&		0&		1&		0&		0&		0&		0&		0\\
		0&		0&		0&		1&		0&		0&		0&		0\\
		0&		0&		0&		0&		1&		0&		0&		0\\
		0&		0&		0&		0&		0&		1&		0&		0\\
		0&		0&		0&		0&		0&		0&		0&		1\\
		0&		0&		0&		0&		0&		0&		1&		0\\
		\end{smallmatrix}\right] 
		$ \\
		& & \\
		CSWAP gate &  $\Qcircuit @C=0.7em @R=1em
		{
			\lstick{\ket{x_1}} & \ctrl{1}  & \rstick{\ket{y_1}}                 \qw \\
			\lstick{\ket{x_2}} & \qswap\qwx  & \rstick{\ket{y_2}}                 \qw \\
			\lstick{\ket{x_3}} & \qswap\qwx      & \rstick{\ket{y_3}} \qw
		}$ &
		$\left[\begin{smallmatrix}1&0&0&0&0&0&0&0\\0&1&0&0&0&0&0&0\\0&0&1&0&0&0&0&0\\0&0&0&1&0&0&0&0\\0&0&0&0&1&0&0&0\\0&0&0&0&0&0&1&0\\0&0&0&0&0&1&0&0\\0&0&0&0&0&0&0&1\\
		\end{smallmatrix}\right]
		$ \\
		& & \\
		\hline
	\end{tabular}
\label{TABLE.1}
\end{table}

\begin{table}[t]
	\caption{Truth tables of CCNOT gate and CSWAP gate}
	\centering  
		\begin{tabular}{cc||cc}
			\hline
			\multicolumn{2}{c||}{CCNOT gate}& \multicolumn{2}{c}{CSWAP gate}\\
			\text{ $\ket{x_1x_2x_3}$} & \text{ $\ket{y_1y_2y_3}$} & \text{ $\ket{x_1x_2x_3}$} & \text{ $\ket{y_1y_2y_3}$} \\
			\hline
			$\ket{000}$ & $\ket{000}$ & $\ket{000}$ & $\ket{000}$  \\
			$\ket{001}$ & $\ket{001}$ & $\ket{001}$ & $\ket{001}$  \\
			$\ket{010}$ & $\ket{010}$ & $\ket{010}$ & $\ket{010}$  \\
			$\ket{011}$ & $\ket{011}$ & $\ket{011}$ & $\ket{011}$  \\
			$\ket{100}$ & $\ket{100}$ & $\ket{100}$ & $\ket{100}$  \\
			$\ket{101}$ & $\ket{101}$ & $\ket{101}$ & $\ket{110}$  \\
			$\ket{110}$ & $\ket{111}$ & $\ket{110}$ & $\ket{101}$  \\
			$\ket{111}$ & $\ket{110}$ & $\ket{111}$ & $\ket{111}$  \\
			\hline
		\end{tabular}
	\label{TABLE.2}
\end{table}

Thirdly, quantum measurements are described by a collection of measurement operators. After  being acted by a certain measurement operator $M_x$, the state of a quantum system $\ket{\psi}$ will collapse into state $\frac{M_x\left| \psi \right>}{\sqrt{\left< \psi \right|M_{x}^{\dag}M_x\left| \psi \right>}}$ with the probability
\begin{equation}
		p( \left| x \right>  ) =\left< \psi  \right|M_{x}^{\dag}M_x\left| \psi  \right>
\end{equation}
where $\dag$ denotes the conjugate transpose.

Fourthly, if a quantum system is composed by several subsystems, then the state space of this composite system  can be represented by
\begin{equation}
	\bigotimes_{i=1}^N{\left| \psi _i \right>}=\left| \psi _1 \right> \otimes \left| \psi _2 \right> \otimes \cdots \otimes \left| \psi _N \right> 
\end{equation}
where $\otimes$ denotes the tensor product, and $\left| \psi _i \right>$ is  the state of the $i$-th component subsystem. For convenience, $\ket{\psi_1} \otimes\ket{\psi_2}$ can be written as  $\ket{\psi_1}\ket{\psi_2}$ or $\ket{\psi_1 \psi_2}$. If  the state of a certain system has the property that $\ket{\psi} \ne \ket{x} \ket{y}$ for all the single qubit states $\ket{x}$  and $\ket{y}$, such as $\frac{1}{\sqrt{2}} (\ket{00} + \ket{11})$, then this special state is called the entangled state.

\subsection{Quantum fuzzy sets}
Given a universe of discourse $X=\{x_j |j = 1,\,2,\, ... \,,\,N\}$, classical fuzzy membership functions $f_i: X \rightarrow [0,1]$ $(i = 1,\,2,\, ... \,,\,k)$ defined on $X$, and $c_i \in \mathbb{C}$ $(i = 1,\,2,\, ... \,,\,k)$, a quantum fuzzy set (QFS) is defined as a linear combination of $N$-dimensional quantum state, given by \cite{mannucci2006quantum}:

\noindent\textbf{Definition 2.6:}
\emph{ Quantum fuzzy sets}

\emph{
\begin{equation}
		\left| s \right> =\sum_{i=1}^k{c_i\left| s_{f_i} \right>}
\end{equation}
where $\left| s_{f_i} \right> =\bigotimes_{j=1}^N{\left( \sqrt{1-f_i\left( x_j \right)}\left| 0 \right> +\sqrt{f_i\left( x_j \right)}\left| 1 \right> \right)}$ is an $N$-dimensional quantum state.
}

\subsection{Quantum fidelity}

In quantum information theory, quantum fidelity is measure of distance between quantum states \cite{nielsen2002quantum}. Given two quantum states represented by density matrices $\rho$ and $\sigma$, the quantum fidelity of these two states is defined as \cite{nielsen2002quantum}:

\noindent\textbf{Definition 2.7:}
\emph{ Quantum fidelity }

\emph{ 
\begin{equation}
		F(\rho,\sigma)=\left(\operatorname{tr}\sqrt{\sqrt{\rho} \sigma \sqrt{\rho}}\right)^2
\end{equation}
where $\operatorname{tr}$ denotes the trace of square matrix.
}

If the two states are pure state $\ket{\psi}$ and $\ket{\phi}$, which means that their associated density matrices are $\rho=\ket{\psi}\bra{\psi}$ and $\sigma=\ket{\phi}\bra{\phi}$, then their corresponding quantum fidelity is $F(\rho,\sigma) = |\left< \psi|\phi \right> |^2	$.

\begin{figure}[h]
	\centering
	\includegraphics[scale=0.33]{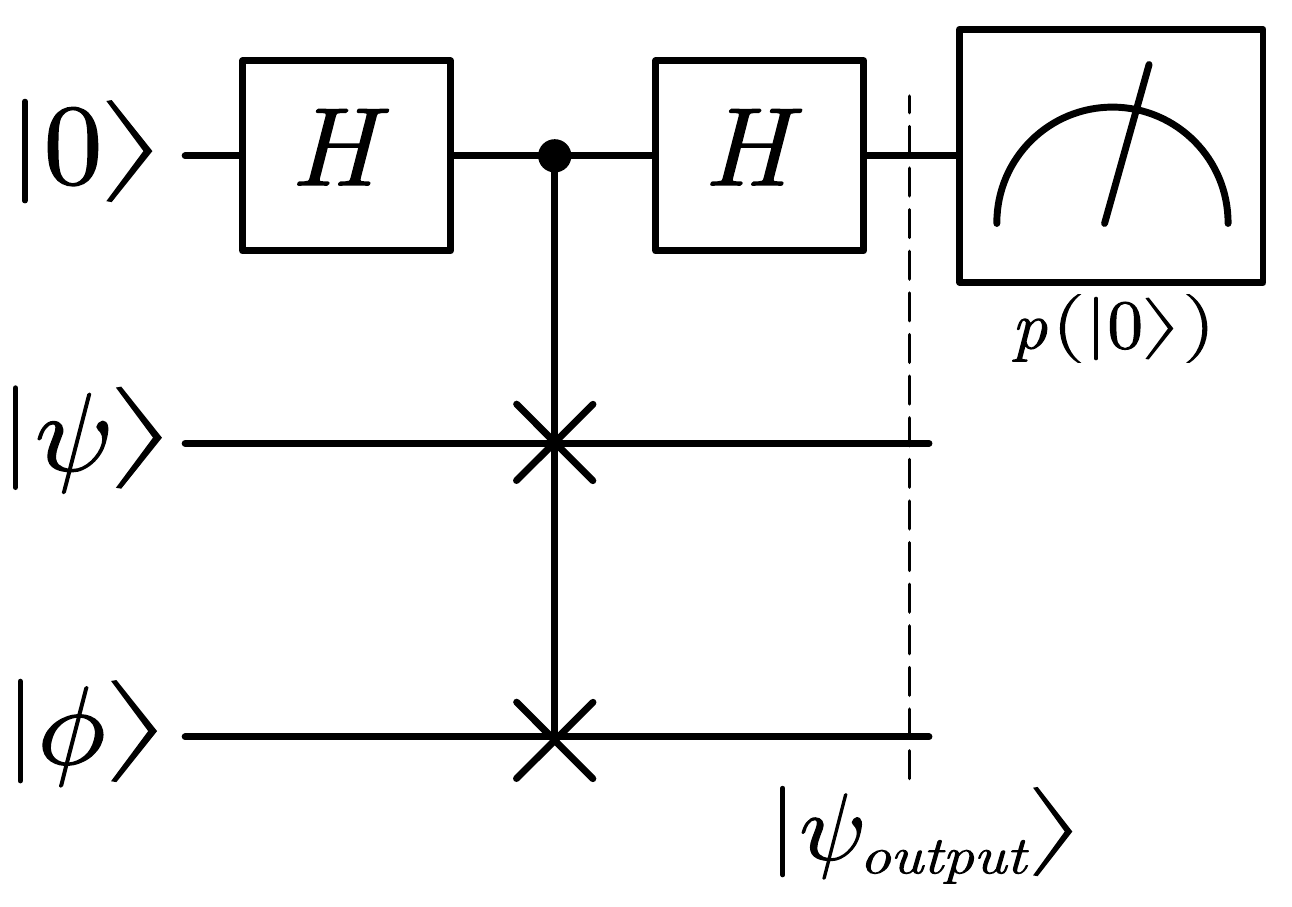}
	\caption{Quantum circuit for implementing  quantum fidelity}
	\label{Fig1}
\end{figure}

For  measuring the similarity between two quantum states, Buhrman proposed a quantum computation trick called swap test, which can calculate the quantum fidelity of two quantum state \cite{buhrman2001quantum}. Its quantum circuit is shown as Fig. \ref{Fig1}.

The input state is $\ket{0}\ket{\psi}\ket{\phi}$ where $\ket{0}$ is the ancillary qubit. Firstly, the ancillary qubit gets through a Hadamard gate. Then, three qubits get through a CSWAP gate. If the ancillary qubit is $\ket{1}$, CSWAP gate will swap $\ket{\psi}\ket{\phi}$  into $\ket{\phi}\ket{\psi}$. Next, the ancillary qubit gets through another Hadamard gate. Finally,  the output state is that
\begin{align}
	\ket{\psi_{output}}=&\left( H\otimes I\otimes I \right) \left( CSWAP \right) \left( H\otimes I\otimes I \right) (\ket{0}\ket{\psi}\ket{\phi})
	\notag\\
	=&\frac{1}{2}\ket{0}(\ket{\psi}\ket{\phi}+\ket{\phi}\ket{\psi})+\frac{1}{2}\ket{1}(\ket{\psi}\ket{\phi}-\ket{\phi}\ket{\psi})
\end{align}
An observation over the ancillary qubit will obtain the outcome $\ket{0}$ with the probability:
$
p\left( \left| 0 \right> \right) =\frac{1}{2}+\frac{1}{2}|\left< \psi |\phi \right> |^2
$
where $|\left< \psi |\phi \right> |^2$ is the quantum fidelity of  $\ket{\psi}$ and $\ket{\phi}$.


\section{QZNs: Quantum Z-numbers}

\subsection{The definition of quantum Z-numbers}
Let $\mathbb{U}$ be a nonempty subset of the Hilbert space $\mathcal{H}$, which is called the quantum universe of discourse (QUOD). Given a quantum states $\left| \varphi \right> \in \mathbb{U}$, quantum Z-numbers (QZNs) are defined as follows:

\noindent\textbf{Definition 3.1:}
\emph{ Quantum Z-numbers }

\emph{
	\begin{equation}
	Z^Q=(A^Q,B^Q)
	\end{equation}
	which consist of two quantum sets:
	\begin{align}
	A^Q&=\{\left. <\left| \varphi \right> ,\left| \psi _{\mu _A} \right> >\,\, \right|\,\left| \varphi \right> \in \mathbb{U}\}
	\\
	B^Q&=\{\left. <\left| \varphi \right> ,\left| \psi _{\mu _B} \right> >\,\, \right|\,\left| \varphi \right> \in \mathbb{U}\}
	\end{align}
	Set $A^Q$ is a quantum fuzzy restriction of a certain quantum state $\left| \varphi \right> \in \mathbb{U}$, and set $B^Q$ is a quantum measure of the reliability of $A^Q$, where $\left| \psi _{\mu _A} \right> :\mathbb{U}\rightarrow \{ \left| \psi \right> | \left| \psi \right>\in \mathcal{H},  \left< \psi | \psi \right> = 1 \} $ and $\left| \psi _{\mu _B} \right> :\mathbb{U}\rightarrow \{ \left| \psi \right> | \left| \psi \right>\in \mathcal{H},  \left< \psi | \psi \right> = 1 \}  $ are respectively the quantum membership function (QMF) of $A^Q$ and $B^Q$, defined as the vectors in a two-dimensional complex Hilbert space:
	\begin{align}
	\left| \psi _{\mu _A} \right> &=\alpha _{\mu _A}\left( \left| \varphi \right> \right) \left| 0 \right> +\beta _{\overline{\mu _A}}\left( \left| \varphi \right> \right) \left| 1 \right> \,\, \left( QMF\,\,of\,\,A^Q \right) 
	\\
	\left| \psi _{\mu _B} \right> &=\alpha _{\mu _B}\left( \left| \varphi \right> \right) \left| 0 \right> +\beta _{\overline{\mu _B}}\left( \left| \varphi \right> \right) \left| 1 \right> \,\, \left( QMF\,\,of\,\,B^Q \right) 
	\end{align}
	where $\alpha _{\mu _A}\left( \left| \varphi \right> \right), \beta _{\overline{\mu _A}}\left( \left| \varphi \right> \right), \alpha _{\mu _B}\left( \left| \varphi \right> \right), \beta _{\overline{\mu _B}}\left( \left| \varphi \right> \right) \in \mathbb{C}$  are respectively the probability amplitude of QMF $\left| \psi _{\mu _A} \right>$ and $\left| \psi _{\mu _B} \right>$ for the state $\ket{0}$ and $\ket{1}$, which are	 constrained by the following conditions:
	\begin{align}
	\left< \psi _{\mu _A}|\psi _{\mu _A} \right> &=\left| \alpha _{\mu _A}\left( \left| \varphi \right> \right) \right|^2+\left| \beta _{\overline{\mu _A}}\left( \left| \varphi \right> \right) \right|^2=1
	\\
	\left< \psi _{\mu _B}|\psi _{\mu _B} \right> &=\left| \alpha _{\mu _B}\left( \left| \varphi \right> \right) \right|^2+\left| \beta _{\overline{\mu _B}}\left( \left| \varphi \right> \right) \right|^2=1
	\end{align}
}

For the convenience of expression, a QZN can be represented as:
\begin{equation}
Z^Q = \left( A^Q,B^Q \right) =\{\left. <\left| \varphi \right> ,\left| \psi _{\mu _A} \right> ,\left| \psi _{\mu _B} \right> >\,\, \right|\,\left| \varphi \right> \in \mathbb{U}\}
\end{equation}

If there is only one element in the QUOD or we just focus on the QZN of one certain element in the QUOD,  the QZN can be written as:
\begin{equation}
Z^Q = < \left| \psi _{\mu _A} \right> ,\left| \psi _{\mu _B} \right> >
\end{equation}

\subsection{Basic quantum fuzzy operations of QZNs}

In this subsection, some basic operations of QZNs are introduced. Let $CCNOT \left( \left| x y z \right> \right)$, $X(\left| x \right>)$, and $R_Y^ \theta  (\left| x \right>) $ respectively denote the CCNOT gate, Pauli-X gate, and Y-Rotation gate in quantum computation. For two QZNs $Z_1^Q$ and $Z_2^Q$ defined on QUOD $\mathbb{U}$, the basic operations of QZNs are defined as follows:

\noindent\textbf{Definition 3.2:}
\emph{ Basic operations of QZNs}

\noindent\emph{(1) \textbf{Inclusion relation:} }

\emph{		
		$Z_1^Q \subseteq Z_2^Q$ if and only if $p( \left| 0 \right> _{\mu _{A1}} ) \le p( \left| 0 \right> _{\mu _{A2}} ) $ and $ p( \left| 0 \right> _{\mu _{B1}} ) \le p( \left| 0 \right> _{\mu _{B2}} ) $, where $p( \left| 0 \right> _{\mu _{k}} )$ is probability of QMF  $\ket{\psi_{\mu _{k}}}$ collapsing into state $\ket{0}$ by observation $(k \in \{A1, A2, B1, B2\})$, which is defined as
		\begin{align}
		p( \left| 0 \right> _{\mu _{A1}} ) =\left< \psi _{\mu _{A1}} \right|M_{0}^{\dag}M_0\left| \psi _{\mu _{A1}} \right> =|\alpha _{\mu _{A1}}\left( \left| \varphi \right> \right) |^2
		\\
		p( \left| 0 \right> _{\mu _{A2}} ) =\left< \psi _{\mu _{A2}} \right|M_{0}^{\dag}M_0\left| \psi _{\mu _{A2}} \right> =|\alpha _{\mu _{A2}}\left( \left| \varphi \right> \right) |^2
		\\
		p( \left| 0 \right> _{\mu _{B1}} ) =\left< \psi _{\mu _{B1}} \right|M_{0}^{\dag}M_0\left| \psi _{\mu _{B1}} \right> =|\alpha _{\mu _{B1}}\left( \left| \varphi \right> \right) |^2
		\\
		p( \left| 0 \right> _{\mu _{B2}} ) =\left< \psi _{\mu _{B2}} \right|M_{0}^{\dag}M_0\left| \psi _{\mu _{B2}} \right> =|\alpha _{\mu _{B2}}\left( \left| \varphi \right> \right) |^2
		\end{align}
		in which $M_0=\left| 0 \right> \left< 0 \right|$ is the measurement operator, and $M_{0}^{\dag}$ means the conjugate transpose of $M_0$.
}
		
\noindent\emph{(2) \textbf{Equality:}} 

\emph{		
		$Z_1^Q = Z_2^Q$ if and only if $Z_1^Q \subseteq  Z_2^Q$ and $Z_2^Q \subseteq  Z_1^Q$.
}
		
\noindent\emph{(3) \textbf{Complement:}}

\emph{
		$\overline{Z^Q}=\{ <\left| \varphi \right> ,\left| \psi _{\mu _A}^C \right> ,\left| \psi _{\mu _B}^C \right> > |\,\,\left| \varphi \right> \in \mathbb{U}\}$, where $\left| \psi _{\mu _A}^C \right>$ and $ \left| \psi _{\mu _B}^C \right>$ are called the complement states of $Z^Q$, which are defined as:
		\begin{align}
		\left| \psi _{\mu _A}^C \right>\triangleq X\left( \left| \psi _{\mu _A} \right> \right) =\beta _{\overline{\mu _A}}\left( \left| \varphi \right> \right) \left| 0 \right> +\alpha _{\mu _A}\left( \left| \varphi \right> \right) \left| 1 \right> 
		\\
		\left| \psi _{\mu _B}^C \right>\triangleq X\left( \left| \psi _{\mu _B} \right> \right) =\beta _{\overline{\mu _B}}\left( \left| \varphi \right> \right) \left| 0 \right> +\alpha _{\mu _B}\left( \left| \varphi \right> \right) \left| 1 \right> 
		\end{align}
}
		
\noindent\emph{(4) \textbf{Intersection:} }

\emph{		
		$Z_{1}^{Q}\cap Z_{2}^{Q}=\{\left. <\left| \varphi \right> ,\left| \psi^I _{\mu_{A12}} \right> ,\left| \psi^I _{\mu _{B12}} \right>  >\,\, \right|\,\left| \varphi \right> \in \mathbb{U}\}$, where $\left| \psi^I _{\mu_{A12}} \right>$ and $ \left| \psi^I _{\mu _{B12}} \right>$ are called the intersection states of $Z_1^Q$ and $Z_2^Q$, which are defined as:
		\begin{align}
		&\left| \psi^I _{\mu_{A12}} \right> \triangleq  CCNOT\left( \left( X\otimes X\otimes I \right) \left| \psi _{\mu _{A1}} \right> \left| \psi _{\mu _{A2}} \right> \left| 1 \right> \right) 
		\\
		&=\alpha _{\mu _{A1}}\left( \left| \varphi \right> \right) \alpha _{\mu _{A2}}\left( \left| \varphi \right> \right) \left| 110 \right> +\alpha _{\mu _{A1}}\left( \left| \varphi \right> \right) \beta _{\overline{\mu _{A2}}}\left( \left| \varphi \right> \right) \left| 101 \right> 
		\notag\\
		&+\beta _{\overline{\mu _{A1}}}\left( \left| \varphi \right> \right) \alpha _{\mu _{A2}}\left( \left| \varphi \right> \right) \left| 011 \right> +\beta _{\overline{\mu _{A1}}}\left( \left| \varphi \right> \right) \beta _{\overline{\mu _{A2}}}\left( \left| \varphi \right> \right) \left| 001 \right> 
		\notag\\
		&\left| \psi^I _{\mu_{B12}} \right> \triangleq  CCNOT\left( \left( X\otimes X\otimes I \right) \left| \psi _{\mu _{B1}} \right> \left| \psi _{\mu _{B2}} \right> \left| 1 \right> \right) 
		\\
		&=\alpha _{\mu _{B1}}\left( \left| \varphi \right> \right) \alpha _{\mu _{B2}}\left( \left| \varphi \right> \right) \left| 110 \right> +\alpha _{\mu _{B1}}\left( \left| \varphi \right> \right) \beta _{\overline{\mu _{B2}}}\left( \left| \varphi \right> \right) \left| 101 \right> 
		\notag\\
		&+\beta _{\overline{\mu _{B1}}}\left( \left| \varphi \right> \right) \alpha _{\mu _{B2}}\left( \left| \varphi \right> \right) \left| 011 \right> +\beta _{\overline{\mu _{B1}}}\left( \left| \varphi \right> \right) \beta _{\overline{\mu _{B2}}}\left( \left| \varphi \right> \right) \left| 001 \right>   \notag
		\end{align}
}		

\noindent\emph{(5) \textbf{Union:} }

\emph{		
		$Z_{1}^{Q}\cup Z_{2}^{Q}=\{\left. <\left| \varphi \right> ,\left| \psi^U _{\mu_{A12}} \right> ,\left| \psi^U _{\mu_{B12}} \right> >\,\, \right|\,\left| \varphi \right> \in \mathbb{U}\}$, where $\left| \psi^U _{\mu_{A12}} \right>$ and $\left| \psi^U _{\mu_{B12}} \right>$ are called the union states of $Z_1^Q$ and $Z_2^Q$, which are defined as:
		\begin{align}
		&\left| \psi^U _{\mu_{A12}} \right> \triangleq CCNOT\left( \left| \psi _{\mu _{A1}} \right> \left| \psi _{\mu _{A2}} \right> \left| 0 \right> \right) 
		\\
		&=\alpha _{\mu _{A1}}\left( \left| \varphi \right> \right) \alpha _{\mu _{A2}}\left( \left| \varphi \right> \right) \left| 000 \right> +\alpha _{\mu _{A1}}\left( \left| \varphi \right> \right) \beta _{\overline{\mu _{A2}}}\left( \left| \varphi \right> \right) \left| 010 \right> 
		\notag\\
		&+\beta _{\overline{\mu _{A1}}}\left( \left| \varphi \right> \right) \alpha _{\mu _{A2}}\left( \left| \varphi \right> \right) \left| 100 \right> +\beta _{\overline{\mu _{A1}}}\left( \left| \varphi \right> \right) \beta _{\overline{\mu _{A2}}}\left( \left| \varphi \right> \right) \left| 111 \right> 
		\notag\\
		&\left| \psi^U _{\mu_{B12}} \right> \triangleq CCNOT\left( \left| \psi _{\mu _{B1}} \right> \left| \psi _{\mu _{B2}} \right> \left| 0 \right> \right) 
		\\
		&=\alpha _{\mu _{B1}}\left( \left| \varphi \right> \right) \alpha _{\mu _{B2}}\left( \left| \varphi \right> \right) \left| 000 \right> +\alpha _{\mu _{B1}}\left( \left| \varphi \right> \right) \beta _{\overline{\mu _{B2}}}\left( \left| \varphi \right> \right) \left| 010 \right> 
		\notag\\
		&+\beta _{\overline{\mu _{B1}}}\left( \left| \varphi \right> \right) \alpha _{\mu _{B2}}\left( \left| \varphi \right> \right) \left| 100 \right> +\beta _{\overline{\mu _{B1}}}\left( \left| \varphi \right> \right) \beta _{\overline{\mu _{B2}}}\left( \left| \varphi \right> \right) \left| 111 \right>   \notag
		\end{align}
}
		
\noindent\emph{(6) \textbf{Convert Z-number into QZN:}}

\emph{		
		Given a classical Z-number $Z=(A,B)$ with the classical membership function $\mu_A$ and $\mu_B$, its corresponding QZN is defined as: $	Z^Q=\{\left. <\left| \varphi \right> ,\left| \psi _{\mu _A} \right> ,\left| \psi _{\mu _B} \right> >\,\, \right|\,\left| \varphi \right> \in \mathbb{U}\} $, where $ \left| \psi _{\mu _A} \right> $ and $	\left| \psi _{\mu _B} \right>  $ are the QMF of QZN. By preparing qubit $\ket{0}$ and passing it through Y-Rotation gate, the QMF of QZN can be obtained:
		\begin{align}
		\left| \psi _{\mu_A} \right> \triangleq R_{Y}^{\theta _{\mu _A}}\left( \left| 0 \right> \right) =\sqrt{\mu _A}\left| 0 \right> +\sqrt{1-\mu _A}\left| 1 \right> 
		\\
		\left| \psi _{\mu_B} \right> \triangleq R_{Y}^{\theta _{\mu _B}}\left( \left| 0 \right> \right) =\sqrt{\mu _B}\left| 0 \right> +\sqrt{1-\mu _B}\left| 1 \right> 
		\end{align}
		where  $\theta _{\mu _A}=2\arccos \left( \sqrt{\mu _A} \right) $ and $ \theta _{\mu _B}=2\arccos \left( \sqrt{\mu _B} \right) $ are generated by the membership function of classical Z-numbers.
}
		
\noindent\emph{(7) \textbf{Combination of QZN:}}
		
\emph{		
	    The combination of a certain QZN $Z^Q = (A^Q, B^Q)$ combines $A^Q$ and $B^Q$, and yields c-QZN defined as: $CZ^Q=\{\left. <\left| \varphi \right> ,\left| \psi _{\mu _{AB}} \right> >\,\, \right|\,\left| \varphi \right> \in \mathbb{U}\}$, where $\left| \psi _{\mu _{AB}} \right>$ is called the combined state of $Z^Q$, which is defined as:
		\begin{align}
		&\left| \psi _{\mu _{AB}} \right> \triangleq  CCNOT\left( \left( X\otimes X\otimes I \right) \left| \psi _{\mu _A} \right> \left| \psi _{\mu _B} \right> \left| 1 \right> \right) 
		\\
		&=\alpha _{\mu _A}\left( \left| \varphi \right> \right) \alpha _{\mu _B}\left( \left| \varphi \right> \right) \left| 110 \right> +\alpha _{\mu _A}\left( \left| \varphi \right> \right) \beta _{\overline{\mu _B}}\left( \left| \varphi \right> \right) \left| 101 \right> 
		\notag\\
		&+\beta _{\overline{\mu _A}}\left( \left| \varphi \right> \right) \alpha _{\mu _B}\left( \left| \varphi \right> \right) \left| 011 \right> +\beta _{\overline{\mu _A}}\left( \left| \varphi \right> \right) \beta _{\overline{\mu _B}}\left( \left| \varphi \right> \right) \left| 001 \right> \notag
		\end{align}
}

\begin{figure}[h]
	\centering
	\includegraphics[scale=0.24]{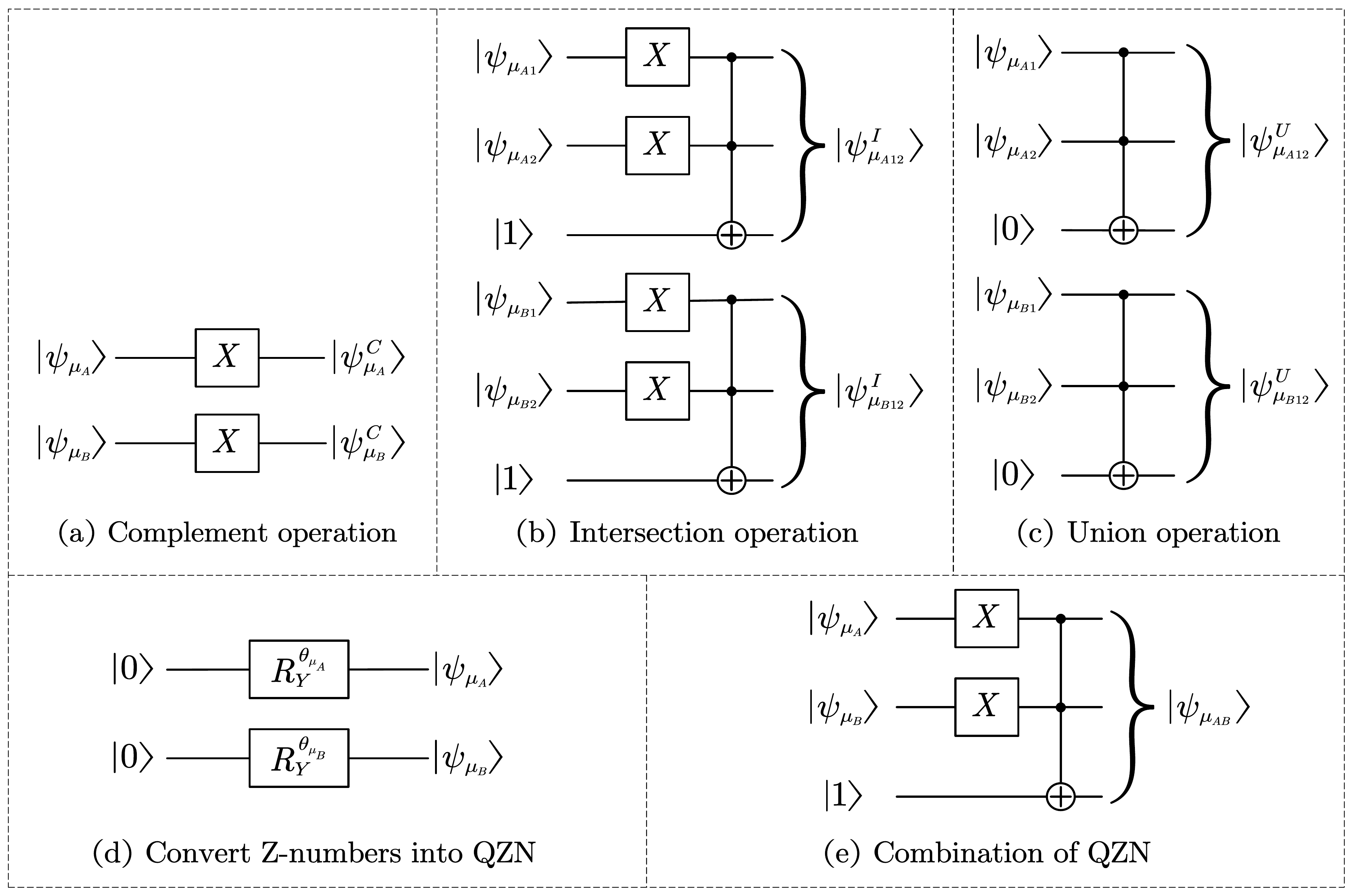}
	\caption{Quantum circuits of the operations (3) to (7)}
	\label{Fig2}
\end{figure}

For implementing the operations of QZNs in quantum computation,  the quantum circuits of the operations (3) to (7) are illustrated in Fig. \ref{Fig2}.

\section{Numerical examples}

In this section, some numerical examples are shown to illustrate QZNs and their operations. After each example, we have a brief discussion. 
In the rest of this section, for convenience, let $k \in \{A1, A2, B1, B2\}$.

\noindent\textbf{Example 4.1:}

Let two QZNs be
\begin{align}
Z_{1}^{Q}=<&\sqrt{0.3}e^{0.7\pi i}\left| 0 \right> +\sqrt{0.7}e^{1.5\pi i}\left| 1 \right> ,
\notag\\
&\sqrt{0.6}e^{0.9\pi i}\left| 0 \right> +\sqrt{0.4}e^{0.3\pi i}\left| 1 \right> >
\\
Z_{2}^{Q}=<&\sqrt{0.4}e^{0.6\pi i}\left| 0 \right> +\sqrt{0.6}e^{0.2\pi i}\left| 1 \right> ,
\notag\\
&\sqrt{0.7}e^{0.5\pi i}\left| 0 \right> +\sqrt{0.3}e^{0.4\pi i}\left| 1 \right> >
\end{align}

According to operation (1) of QZNs,  calculate the probability: 
$p( \left| 0 \right> _{\mu _{A1}} ) =0.3$, 
$p( \left| 0 \right> _{\mu _{A2}} ) =0.4$, 
$p( \left| 0 \right> _{\mu _{B1}} ) =0.6$, 
$p( \left| 0 \right> _{\mu _{B2}} ) =0.7$.
Because of $p( \left| 0 \right> _{\mu _{A1}} ) \le p( \left| 0 \right> _{\mu _{A2}} ) $ and $  p( \left| 0 \right> _{\mu _{B1}} ) \le p( \left| 0 \right> _{\mu _{B2}} )$, the inclusion relation of these two QZNs is $Z_{1}^{Q}\subseteq Z_{2}^{Q}$.

This example shows that the inclusion relation of QZNs compares the probability $p( \left| 0 \right> _{\mu _{k}} )$ of different QMF $\ket{\psi_{\mu _{k}}}$ collapsing into state $\ket{0}$ by observation. The larger the magnitude of the probability amplitude $|\alpha_{\mu _{k}}|$ is, the higher the corresponding probability $p( \left| 0 \right> _{\mu _{k}} )$ is.

\noindent\textbf{Example 4.2:}

Let two QZNs be
\begin{align}
Z_{1}^{Q}=<&\sqrt{0.3}e^{0.7\pi i}\left| 0 \right> +\sqrt{0.7}e^{1.5\pi i}\left| 1 \right> ,
\notag\\
&\sqrt{0.6}e^{0.9\pi i}\left| 0 \right> +\sqrt{0.4}e^{0.3\pi i}\left| 1 \right> >
\\
Z_{2}^{Q}=<&\sqrt{0.3}e^{0.6\pi i}\left| 0 \right> +\sqrt{0.7}e^{0.2\pi i}\left| 1 \right> ,
\notag\\
&\sqrt{0.6}e^{0.5\pi i}\left| 0 \right> +\sqrt{0.4}e^{0.4\pi i}\left| 1 \right> >
\end{align}

Based on operation (1) of QZNs, it can be calculated that 
$p( \left| 0 \right> _{\mu _{A1}} ) =0.3$, 
$p( \left| 0 \right> _{\mu _{A2}} ) =0.3$, 
$p( \left| 0 \right> _{\mu _{B1}} ) =0.6$, 
$p( \left| 0 \right> _{\mu _{B2}} ) =0.6$.
Next,  $p( \left| 0 \right> _{\mu _{A1}} ) \le p( \left| 0 \right> _{\mu _{A2}} ) $ and $  p( \left| 0 \right> _{\mu _{B1}} ) \le p( \left| 0 \right> _{\mu _{B2}} )$, so $Z_{1}^{Q}\subseteq Z_{2}^{Q}$. Also, $Z_{2}^{Q}\subseteq Z_{1}^{Q}$ since $p( \left| 0 \right> _{\mu _{A2}} ) \le p( \left| 0 \right> _{\mu _{A1}} ) $ and $  p( \left| 0 \right> _{\mu _{B2}} ) \le p( \left| 0 \right> _{\mu _{B1}} )$. As a result, according to operation (2) of QZNs, we can draw the conclusion that $Z_{1}^{Q} = Z_{2}^{Q}$.

This example shows that the equality of QZNs compares the probability $p( \left| 0 \right> _{\mu _{k}} )$ of different QMF  $\ket{\psi_{\mu _{k}}}$ collapsing into state $\ket{0}$ by observation. If and only if $p( \left| 0 \right> _{\mu _{A1}} ) = p( \left| 0 \right> _{\mu _{A2}} ) $ and $  p( \left| 0 \right> _{\mu _{B1}} ) = p( \left| 0 \right> _{\mu _{B2}} )$, two QZNs are equal.

\noindent\textbf{Example 4.3:}

Let a QZN be $Z^Q=<\sqrt{0.3}\left| 0 \right> +\sqrt{0.7}\left| 1 \right> ,\sqrt{0.6}\left| 0 \right> +\sqrt{0.4}\left| 1 \right> >$, which is converted by a classical Z-number $Z_1 = <0.3, 0.6>$. Based on the operation (3) of QZN, the complement state of this QZN can be calculated as:
\begin{align}
&\left| \psi _{\mu _A}^{C} \right> =X\left( \sqrt{0.3}\left| 0 \right> +\sqrt{0.7}\left| 1 \right> \right) =\sqrt{0.7}\left| 0 \right> +\sqrt{0.3}\left| 1 \right> 
\\
&\left| \psi _{\mu _B}^{C} \right> =X\left( \sqrt{0.6}\left| 0 \right> +\sqrt{0.4}\left| 1 \right> \right) =\sqrt{0.4}\left| 0 \right> +\sqrt{0.6}\left| 1 \right> 
\end{align}
And then, the complement of this QZN can be obtained: $\overline{Z^Q}=<\sqrt{0.7}\left| 0 \right> +\sqrt{0.3}\left| 1 \right> ,\sqrt{0.4}\left| 0 \right> +\sqrt{0.6}\left| 1 \right> >$.

When an observation is performed over $\left| \psi _{\mu _A}^{C} \right>$ and $\left| \psi _{\mu _B}^{C} \right>$, they will collapse into state $\ket{0}$ with the probability:
\begin{align}
p( \left| 0 \right> _{\mu _A} ) =\left< \psi _{\mu _A}^{C} \right|M_{0}^{\dag}M_0\left| \psi _{\mu _A}^{C} \right> = 0.7
\\
p( \left| 0 \right> _{\mu _B} ) =\left< \psi _{\mu _B}^{C} \right|M_{0}^{\dag}M_0\left| \psi _{\mu _B}^{C} \right> = 0.4
\end{align}
where $M_0=\left| 0 \right> \left< 0 \right|$ is the measurement operator. These probabilities can be composed into a classical Z-number $Z_2=<0.7,0.4>$, which is the complement of $Z_1 = <0.3, 0.6>$ based on the fuzzy complement operation \cite{klir1995fuzzy}.

This example shows that the complement of QZN can be implemented by applying Pauli-X gate. In addition, the complement of QZN is compatible with the classical complement operation of Z-number. The  complement of QZN can degenerate into classical complement operation of Z-number by performing an observation of the complement state and obtaining their probability of collapsing into state $\ket{0}$.

\noindent\textbf{Example 4.4:}

Let two QZNs be 
\begin{align}
Z_{1}^{Q}=<\sqrt{x_1}\left| 0 \right> +\sqrt{1-x_1}\left| 1 \right> ,\sqrt{x_2}\left| 0 \right> +\sqrt{1-x_2}\left| 1 \right> >
\\
Z_{2}^{Q}=<\sqrt{y_1}\left| 0 \right> +\sqrt{1-y_1}\left| 1 \right> ,\sqrt{y_2}\left| 0 \right> +\sqrt{1-y_2}\left| 1 \right> >
\end{align}
which are converted by two classical Z-numbers $Z_1=<x_1,x_2>$ and $Z_2=<y_1,y_2>$. Based on the operation (4) of QZNs, the intersection states  of these two QZNs are that
\begin{align}
\left| \psi _{\mu _{A12}}^{I} \right> =&CCNOT\left( \left( X\otimes X\otimes I \right) \left| \psi _{\mu _{A1}} \right> \left| \psi _{\mu _{A2}} \right> \left| 1 \right> \right) 
\\
=&\sqrt{x_1y_1}\left| 110 \right> +\sqrt{x_1\left( 1-y_1 \right)}\left| 101 \right> 
\notag\\
&+\sqrt{\left( 1-x_1 \right) y_1}\left| 011 \right> +\sqrt{\left( 1-x_1 \right) \left( 1-y_1 \right)}\left| 001 \right> 
\notag\\
\left| \psi _{\mu _{B12}}^{I} \right> =&CCNOT\left( \left( X\otimes X\otimes I \right) \left| \psi _{\mu _{B1}} \right> \left| \psi _{\mu _{B2}} \right> \left| 1 \right> \right) 
\\
=&\sqrt{x_2y_2}\left| 110 \right> +\sqrt{x_2\left( 1-y_2 \right)}\left| 101 \right> 
\notag\\
&+\sqrt{\left( 1-x_2 \right) y_2}\left| 011 \right> +\sqrt{\left( 1-x_2 \right) \left( 1-y_2 \right)}\left| 001 \right> \notag
\end{align}
Hence, the intersection of $Z_{1}^{Q}$ and $Z_{2}^{Q}$ is that $
Z_{1}^{Q}\cap Z_{2}^{Q}=<\left| \psi _{\mu _{A12}}^{I} \right>,\left| \psi _{\mu _{B12}}^{I} \right> >$. 

By performing an observation over the third qubit, $\left| \psi _{\mu _{A12}}^{I} \right>$ and $\left| \psi _{\mu _{B12}}^{I} \right>$ will collapse into state $\ket{0}$ with the probability:
\begin{align}
p( \left| 0 \right> _{\mu _{A12}}^{I} ) =\left< \psi _{\mu _{A12}}^{I} \right|M_{0}^{\left( 3 \right) \dag}M_{0}^{\left( 3 \right)}\left| \psi _{\mu _{A12}}^{I} \right> =x_1y_1
\\
p( \left| 0 \right> _{\mu _{B12}}^{I} ) =\left< \psi _{\mu _{B12}}^{I} \right|M_{0}^{\left( 3 \right) \dag}M_{0}^{\left( 3 \right)}\left| \psi _{\mu _{B12}}^{I} \right> =x_2y_2
\end{align}
where $M_{0}^{\left( 3 \right)}=I\otimes I\otimes \left| 0 \right> \left< 0 \right|$ is the measurement operator. These probabilities can be constructed into a classical Z-number $Z_3=<x_1y_1,x_2y_2>$.

\begin{figure}[h]
	\centering
	\includegraphics[scale=0.55]{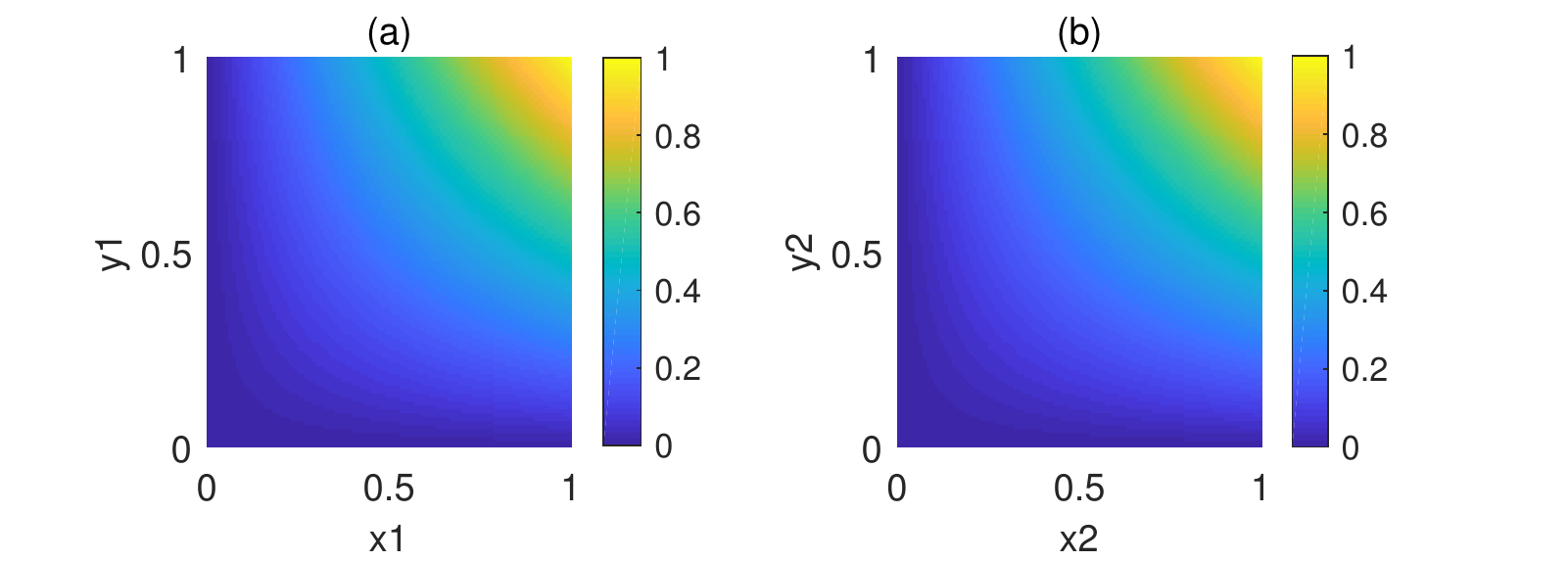}
	\caption{(a) The relationship of $x_1$, $y_1$, $x_1y_1$;\\ (b) The relationship of $x_2$, $y_2$, $x_2y_2$.}
	\label{fig_exp1}
\end{figure}

According to the t-norm operation \cite{klir1995fuzzy}, it can be found that the first component for $Z_3$, namely $x_1y_1$, is the t-norm of the first component for $Z_1$ and $Z_2$, namely $x_1$ and $y_1$;  while the second component for $Z_3$, namely $x_2y_2$, is the t-norm of the second component for $Z_1$ and $Z_2$, namely $x_2$ and $y_2$. To illustrate the intersection of QZNs and t-norm of Z-numbers, the relationship of $x_1$, $y_1$, $x_1y_1$, and the relationship of $x_2$, $y_2$, $x_2y_2$ are shown in Fig. \ref{fig_exp1}.

This example shows that  the intersection of QZNs can be obtain based on CCNOT gate and Pauli-X gate. Next, the intersection of QZNs is compatible with the classical t-norm of Z-numbers \cite{klir1995fuzzy}. The intersection of QZNs can degenerate into t-norm of Z-numbers by performing an observation over the third qubit of the intersection states and obtaining their probabilities of collapsing into state $\ket{0}$.

\noindent\textbf{Example 4.5:}

Let two QZNs be 
\begin{align}
Z_{1}^{Q}=<\sqrt{x_1}\left| 0 \right> +\sqrt{1-x_1}\left| 1 \right> ,\sqrt{x_2}\left| 0 \right> +\sqrt{1-x_2}\left| 1 \right> >
\\
Z_{2}^{Q}=<\sqrt{y_1}\left| 0 \right> +\sqrt{1-y_1}\left| 1 \right> ,\sqrt{y_2}\left| 0 \right> +\sqrt{1-y_2}\left| 1 \right> >
\end{align}
which are converted by two classical Z-numbers $Z_1=<x_1,x_2>$ and $Z_2=<y_1,y_2>$. According to the operation (5) of QZNs, the union states  of these two QZNs are that
\begin{align}
\left| \psi _{\mu _{A12}}^{U} \right> =&CCNOT\left( \left| \psi _{\mu _{A1}} \right> \left| \psi _{\mu _{A2}} \right> \left| 0 \right> \right) 
\\
=&\sqrt{x_1y_1}\left| 000 \right> +\sqrt{x_1\left( 1-y_1 \right)}\left| 010 \right> 
\notag\\
&+\sqrt{\left( 1-x_1 \right) y_1}\left| 100 \right> +\sqrt{\left( 1-x_1 \right) \left( 1-y_1 \right)}\left| 111 \right> 
\notag\\
\left| \psi _{\mu _{A12}}^{U} \right> =&CCNOT\left( \left| \psi _{\mu _{B1}} \right> \left| \psi _{\mu _{B2}} \right> \left| 0 \right> \right) 
\\
=&\sqrt{x_2y_2}\left| 000 \right> +\sqrt{x_2\left( 1-y_2 \right)}\left| 010 \right> 
\notag\\
&+\sqrt{\left( 1-x_2 \right) y_2}\left| 100 \right> +\sqrt{\left( 1-x_2 \right) \left( 1-y_2 \right)}\left| 111 \right>  \notag
\end{align}
As a result, the union of $Z_{1}^{Q}$ and $Z_{2}^{Q}$ is that $
Z_{1}^{Q}\cup Z_{2}^{Q}=<\left| \psi _{\mu _{A12}}^{U} \right>,\left| \psi _{\mu _{B12}}^{U} \right> >$. 

After observing over the third qubit, $\left| \psi _{\mu _{A12}}^{U} \right>$ and $\left| \psi _{\mu _{B12}}^{U} \right>$ collapse into state $\ket{0}$ with the probability:
\begin{align}
&p( \left| 0 \right> _{\mu _{A12}}^{U} ) =\left< \psi _{\mu _{A12}}^{U} \right|M_{0}^{\left( 3 \right) \dag}M_{0}^{\left( 3 \right)}\left| \psi _{\mu _{A12}}^{U} \right> 
\\
&=x_1y_1 + x_1(1-y_1)+(1-x_1)y_1=x_1+y_1-x_1y_1
\notag\\
&p( \left| 0 \right> _{\mu _{B12}}^{U} ) =\left< \psi _{\mu _{B12}}^{U} \right|M_{0}^{\left( 3 \right) \dag}M_{0}^{\left( 3 \right)}\left| \psi _{\mu _{B12}}^{U} \right> 
\\
&=x_2y_2 + x_2(1-y_2)+(1-x_2)y_2=x_2+y_2-x_2y_2
\notag
\end{align}
where $M_{0}^{\left( 3 \right)}=I\otimes I\otimes \left| 0 \right> \left< 0 \right|$ is the measurement operator. These probabilities can be composed into a classical Z-number $Z_3=<x_1+y_1-x_1y_1,x_2+y_2-x_2y_2>$.

\begin{figure}[h]
	\centering
	\includegraphics[scale=0.55]{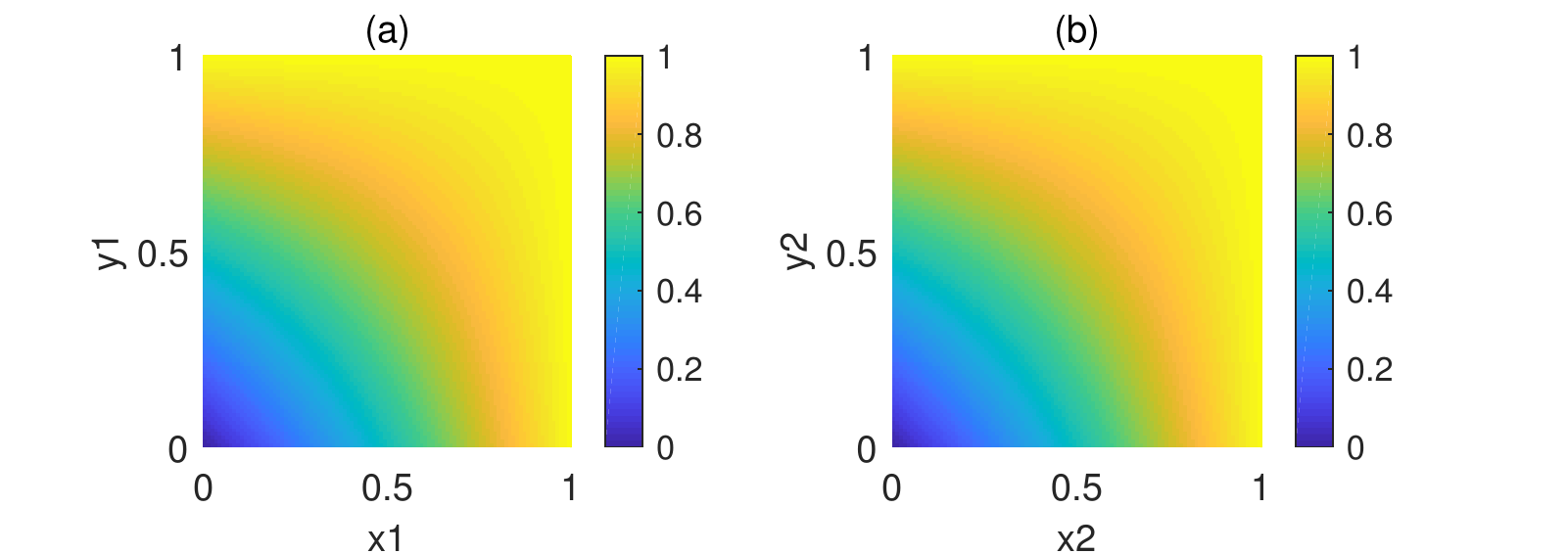}
	\caption{(a) The relationship of $x_1$, $y_1$, $x_1+y_1-x_1y_1$;\\ (b) The relationship of $x_2$, $y_2$, $x_2+y_2-x_2y_2$.}
	\label{fig_exp2}
\end{figure}

Based on the t-conorm operation \cite{klir1995fuzzy}, it can be found that the first component for $Z_3$, namely $x_1+y_1-x_1y_1$, is the t-conorm of the first component for $Z_1$ and $Z_2$, namely $x_1$ and $y_1$;  while the second component for $Z_3$, namely $x_2+y_2-x_2y_2$, is the t-norm of the second component for $Z_1$ and $Z_2$, namely $x_2$ and $y_2$. To illustrate the union of QZNs and t-conorm of Z-numbers, the relationship of $x_1$, $y_1$, $x_1+y_1-x_1y_1$, and the relationship of $x_2$, $y_2$, $x_2+y_2-x_2y_2$ are shown in Fig. \ref{fig_exp2}.

This example shows that  the union of QZNs can be obtain based on CCNOT gate. In addition,  the union of QZNs is compatible with the classical t-conorm of Z-numbers \cite{klir1995fuzzy}. The union of QZNs will degenerate into t-norm of Z-numbers under the condition of an observation over the third qubit of the union states and obtaining their probabilities of collapsing into state $\ket{0}$.

\noindent\textbf{Example 4.6:}

Let a classical Z-number be $Z=<0.5,0.75>$. According to operation (6) of QZN, we can obtain that:
\begin{align}
	&\theta _{\mu _A}=2\text{arc}\cos \left( \sqrt{0.5} \right) =\frac{1}{2}\pi 
	\\
	&\theta _{\mu _B}=2\text{arc}\cos \left( \sqrt{0.75} \right) =\frac{1}{3}\pi 
\end{align}

Based on Y-Rotation gate, we can get the QMF of this Z-number :
\begin{align}
	&\left| \psi _{\mu _A} \right> =R_{Y}^{\theta _{\mu _A}}\left( \left| 0 \right> \right) =\sqrt{0.5}\left| 0 \right> +\sqrt{0.5}\left| 1 \right> 
	\\
	&\left| \psi _{\mu _B} \right> =R_{Y}^{\theta _{\mu _B}}\left( \left| 0 \right> \right) =\sqrt{0.75}\left| 0 \right> +\sqrt{0.25}\left| 1 \right> 
\end{align}

Then the corresponding QZN of this classical Z-number is that $Z^Q=<\sqrt{0.5}\left| 0 \right> +\sqrt{0.5}\left| 1 \right> ,\sqrt{0.75}\left| 0 \right> +\sqrt{0.25}\left| 1 \right> >$.

This example shows that a classical Z-number can be converted into QZN by implementing Y-Rotation gate.

\noindent\textbf{Example 4.7:}

Given a QZN $Z^Q=<\sqrt{x}\left| 0 \right> +\sqrt{1-x}\left| 1 \right> ,\sqrt{y}\left| 0 \right> +\sqrt{1-y}\left| 1 \right> >$ which is converted a classical Z-number $Z=<x,y>$, the combined state  of this QZN can be calculated by operation (7) of QZN:
\begin{align}
\left| \psi _{\mu _{AB}} \right> =&CCNOT\left( \left( X\otimes X\otimes I \right) \left| \psi _{\mu _A} \right> \left| \psi _{\mu _B} \right> \left| 1 \right> \right) 
\\
=&\sqrt{xy}\left| 110 \right> +\sqrt{x\left( 1-y \right)}\left| 101 \right> 
\notag\\
&+\sqrt{\left( 1-x \right) y}\left| 011 \right> +\sqrt{\left( 1-x \right) \left( 1-y \right)}\left| 001 \right>   \notag
\end{align}

Then the c-QZN of this QZN is that $CZ^Q=<\sqrt{xy}\left| 110 \right> +\sqrt{x\left( 1-y \right)}\left| 101 \right> +\sqrt{\left( 1-x \right) y}\left| 011 \right> +\sqrt{\left( 1-x \right) \left( 1-y \right)}\left| 001 \right> >$, which is actually a quantum fuzzy set (QFS) \cite{mannucci2006quantum}. It shows that QZN can degenerate into QFS by applying the combination of QZN. 

An observation over the third qubit makes $\left| \psi _{\mu _{AB}}\right>$ collapsing into state $\ket{0}$ and $\ket{1}$  with the probability:
\begin{align}
&p( \left| 0 \right> _{\mu _{AB}} ) =\left< \psi _{\mu _{AB}} \right|M_{0}^{\left( 3 \right) \dag}M_{0}^{\left( 3 \right)}\left| \psi _{\mu _{AB}} \right> =xy
\\
&p( \left| 1 \right> _{\mu _{AB}} ) =\left< \psi _{\mu _{AB}} \right|M_{1}^{\left( 3 \right) \dag}M_{1}^{\left( 3 \right)}\left| \psi _{\mu _{AB}} \right> =1 - xy
\end{align}
where $M_{0}^{\left( 3 \right)}=I\otimes I\otimes \left| 0 \right> \left< 0 \right|$ and $M_{1}^{\left( 3 \right)}=I\otimes I\otimes \left| 1 \right> \left< 1 \right|$ are the measurement operators. 

\begin{figure}[h]
	\centering
	\includegraphics[scale=0.55]{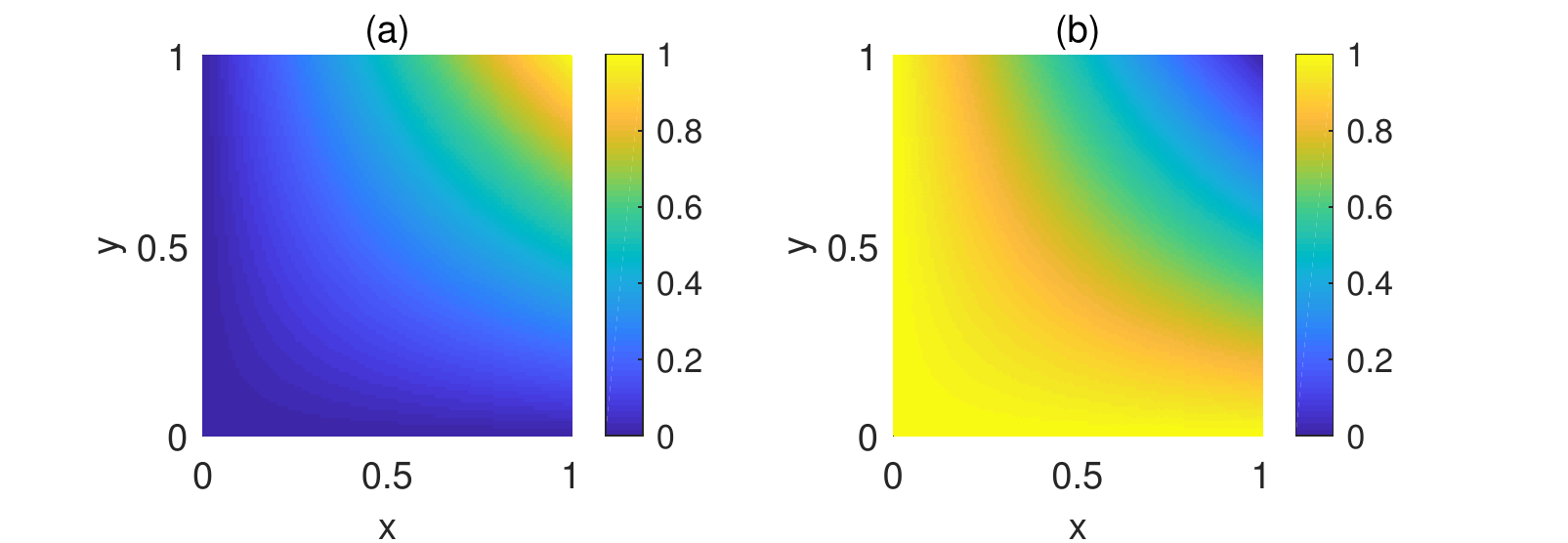}
	\caption{(a) The relationship of $x$, $y$, $xy$;\\ (b) The relationship of $x$, $y$, $1-xy$.}
	\label{fig_exp3}
\end{figure}

Because $x$ and $y$ are the first membership function and the second membership function of $Z=<x,y>$,  $p( \left| 0 \right> _{\mu _{AB}} )=xy$ can be seen as a combination of Z-numbers, and $p( \left| 1 \right> _{\mu _{AB}} )=1-xy$ can be seen as a negation of the combination of Z-numbers. In specific, $xy$ can be interpreted that $x$ is discounted by $y$, and $1-xy$ is the negation of $xy$.   To illustrate,  the relationship of $x$, $y$, $xy$, and the relationship of $x$, $y$, $1-xy$  are shown in Fig. \ref{fig_exp3}.

This example shows that the combination of QZN is based on CCNOT gate and Pauli-X gate. Moreover, the combination of QZN can degenerate into the combination  of Z-number by observing over the third qubit of the combined state and obtaining the probabilities of collapsing into state $\ket{0}$ and  $\ket{1}$.

\section{QZN-based quantum multi-attributes decision making algorithm}

Making decision among different schemes with multi-attributes has attracted many attentions. Researchers have proposed lots of multi-attributes decision making (MADM) algorithm based on various methods, such as soft likelihood functions \cite{fei2019pythagorean}, fuzzy set theory \cite{Xue2020Refined}, evidence theory \cite{deng2020uncertainty}, Z-numbers \cite{aliev2014decision,aliev2016ranking}, D-numbers \cite{mo2020emergency}, distance \cite{wang2017multi}, and divergence \cite{song2019new}. Because of the significance of the theoretical and practical use, MADM algorithms have been applied in risk analysis \cite{mo2021swot,wang2021risk}, quality goals evaluation \cite{mo2020new}, and medical diagnosis \cite{zhou2020new,pan2021constrained}.

In this section,  we propose a novel quantum MADM algorithm based on QZNs. Next, the time complexity of  the proposed algorithm is analyzed.

\subsection{The proposed QZN-based quantum MADM algorithm}

Assume that there are $M$ samples, and each sample has $K$ attributes.   These samples are evaluated by experts or sensors, which are expressed by classical Z-numbers. Assume that there exists $N$ references of these samples, each reference contains $K$ attributes. These references are derived from big data and statistic, which are also expressed by classical Z-numbers. The aim is to classify the $M$ samples and match them to the $N$ references. It should be noted that the number of the attributes of samples is equal to that of references, while the number of samples may not be equal to that of references.

Let $S$ denotes the UOD of all the samples, and $S_{ij}\in S$ denotes the $i$-th sample and its $j$-th attribute, where $i=1, 2, \, … \, , M$ and $j=1, 2, \, … \, , K$. Let $R$ denotes the UOD of all the references, and  $R_{xy}\in R$ denotes the $x$-th reference and its $y$-th attribute, where $x=1, 2, \, … \, , N$ and $y=1, 2, \, … \, , K$. Then, the proposed algorithm proceeds as the following steps.

\noindent\textbf{Step (1): Convert Z-numbers into SZM and RZM.}

	Input the classical Z-numbers of the samples $Z_{S_{ij}}$ and references $Z_{R_{ij}}$, and convert them into sample Z-numbers matrix (SZM) and reference Z-numbers matrix (RZM):
	\begin{align}
	SZM&=\left[ \begin{matrix}
	\overrightarrow{\boldsymbol{Z}_{\boldsymbol{S}_{\boldsymbol{1}}}}&		\overrightarrow{\boldsymbol{Z}_{\boldsymbol{S}_{\boldsymbol{2}}}}&		\cdots&		\overrightarrow{\boldsymbol{Z}_{\boldsymbol{S}_{\boldsymbol{i}}}}&		\cdots&		\overrightarrow{\boldsymbol{Z}_{\boldsymbol{S}_{\boldsymbol{M}}}}\\
	\end{matrix} \right] ^T
	\\
	RZM&=\left[ \begin{matrix}
	\overrightarrow{\boldsymbol{Z}_{\boldsymbol{R}_{\boldsymbol{1}}}}&		\overrightarrow{\boldsymbol{Z}_{\boldsymbol{R}_{\boldsymbol{2}}}}&		\cdots&		\overrightarrow{\boldsymbol{Z}_{\boldsymbol{R}_{\boldsymbol{x}}}}&		\cdots&		\overrightarrow{\boldsymbol{Z}_{\boldsymbol{R}_{\boldsymbol{N}}}}\\
	\end{matrix} \right] ^T
	\end{align}
	in which
	\begin{align}
	\overrightarrow{\boldsymbol{Z}_{\boldsymbol{S}_{\boldsymbol{i}}}}&=\left( \,\,Z_{S_{i1}},\,\,Z_{S_{i2}},\,\,...\,\,,\,\,Z_{S_{ij}},\,\,...\,\,,\,\,Z_{S_{iK}}\,\, \right) \\
	\overrightarrow{\boldsymbol{Z}_{\boldsymbol{R}_{\boldsymbol{x}}}}&=\left( \,\,Z_{R_{x1}},\,\,Z_{R_{x2}},\,\,...\,\,,\,\,Z_{R_{xy}},\,\,...\,\,,\,\,Z_{R_{xK}}\,\, \right) 
	\end{align}  are respectively the sample Z-numbers vector (SZV) and the reference Z-numbers vector (RZV), where $Z_{S_{ij}}=<\mu _A\left( S_{ij} \right) , \mu _B\left( S_{ij} \right) > $ and $\,\,Z_{R_{xy}}=<\mu _A\left( R_{xy} \right) , \mu _B\left( R_{xy} \right) >$ are respectively the Z-number for the $j$-th attribute of the $i$-th sample, and the Z-number for the $y$-th attribute of the $x$-th reference. 
	
\noindent\textbf{Step (2): Calculate the rotation angles of SZM and RZM.}
	
	Calculate the corresponding rotation angles for every element of SZM and RZM, which are defined as:
	\begin{align}
	\theta _{\mu _AS_{ij}}&=2\arccos \left( \sqrt{\mu _A\left( S_{ij} \right)} \right) 
	\\
	\theta _{\mu _BS_{ij}}&=2\arccos \left( \sqrt{\mu _B\left( S_{ij} \right)} \right) 
	\\
	\theta _{\mu _AR_{xy}}&=2\arccos \left( \sqrt{\mu _A\left( R_{xy} \right)} \right) 
	\\
	\theta _{\mu _BR_{xy}}&=2\arccos \left( \sqrt{\mu _B\left( R_{xy} \right)} \right) 
	\end{align}
	where $ i \in \{1,2, ... , M\}  $, $ x \in \{1,2, ... , N\}  $, and $ j,y\in \{1,2, ... , K\}  $.

\noindent\textbf{Step (3): Calculate the similarity of SZV and RZV based on quantum fidelity coefficient.}
	
	For one SZV $\overrightarrow{\boldsymbol{Z}_{\boldsymbol{S}_{\boldsymbol{i}}}}$ and one RZV $\overrightarrow{\boldsymbol{Z}_{\boldsymbol{R}_{\boldsymbol{x}}}}$, from step (3-1) to (3-6), calculate their similarity based on quantum fidelity coefficient $F_{ix}$, until all $F_{ix}$ are calculated, where $ i \in \{1,2, ... , M\}  $ and $ x \in \{1,2, ... , N\}  $.

\noindent\textbf{Step (3-1):}
		Prepare quantum ground state constructed by $4K$ dimensional qubits  $\ket{\phi_0} = \left| 0 \right> ^{\otimes 4K}$.

\noindent\textbf{Step (3-2):}
        According to the operation (6) in \textbf{Definition 3.3}, convert $Z_{S_i}$ and $Z_{R_x}$ into sample QZN vector (SQZV) and reference QZN vector (RQZV):
		\begin{align}
		\overrightarrow{\boldsymbol{Z}_{\boldsymbol{S}_{\boldsymbol{i}}}^{\boldsymbol{Q}}}&=\left( \,\,Z_{S_{i1}}^{Q},\,\,Z_{S_{i2}}^{Q},\,\,...\,\,,\,\,Z_{S_{ij}}^{Q},\,\,...\,\,,\,\,Z_{S_{iK}}^{Q}\,\, \right) 
		\\
		\overrightarrow{\boldsymbol{Z}_{\boldsymbol{R}_{\boldsymbol{x}}}^{\boldsymbol{Q}}}&=\left( \,\,Z_{R_{x1}}^{Q},\,\,Z_{R_{x2}}^{Q},\,\,...\,\,,\,\,Z_{R_{xy}}^{Q},\,\,...\,\,,\,\,Z_{R_{xK}}^{Q}\,\, \right) 
		\end{align}
		in which $Z_{S_{ij}}^{Q} (j = 1, 2, \, ... \, , K)$ and $Z_{R_{xy}}^{Q} (y = 1, 2, \, ... \, , K)$ are respectively the QZNs for the $j$-th attributes of the $i$-th samples and $y$-th attributes of the $x$-th references, which are defined as:
		\begin{align}
		Z_{S_{ij}}^{Q}=<&\left| \psi _{\mu _AS_{ij}} \right> ,\left| \psi _{\mu _BS_{ij}} \right> >\\
		=<&R_{Y}^{\theta _{\mu _AS_{ij}}}\left( \left| 0 \right> \right) ,R_{Y}^{\theta _{\mu _BS_{ij}}}\left( \left| 0 \right> \right) >
		\notag\\
		=<&\sqrt{\mu _A\left( S_{ij} \right)}\left| 0 \right> +\sqrt{1-\mu _A\left( S_{ij} \right)}\left| 1 \right> , 
		\notag\\
		&\sqrt{\mu _B\left( S_{ij} \right)}\left| 0 \right> +\sqrt{1-\mu _B\left( S_{ij} \right)}\left| 1 \right> >
		\notag\\
		Z_{R_{xy}}^{Q}=<&\left| \psi _{\mu _AR_{xy}} \right> ,\left| \psi _{\mu _BR_{xy}} \right> >\\
		=<&R_{Y}^{\theta _{\mu _AR_{xy}}}\left( \left| 0 \right> \right) ,R_{Y}^{\theta _{\mu _BR_{xy}}}\left( \left| 0 \right> \right) >
		\notag\\
		=<&\sqrt{\mu _A\left( R_{xy} \right)}\left| 0 \right> +\sqrt{1-\mu _A\left( R_{xy} \right)}\left| 1 \right> , 
		\notag\\
		&\sqrt{\mu _B\left( R_{xy} \right)}\left| 0 \right> +\sqrt{1-\mu _B\left( R_{xy} \right)}\left| 1 \right> > \notag
		\end{align}
		where $\theta _{\mu _AS_{ij}}$ and $\theta _{\mu _AR_{xy}}$ are respectively the rotation angle for the Y-Rotation gates defined in Step (2).
		
		The way for implementing this step in quantum circuit is to pass $\ket{\phi_0}$ through $4K$ Y-Rotation gates, then the quantum state of SQZV and RQZV  $\ket{\phi_1}$ can be obtained:
		\begin{align}
		\left| \phi _1 \right> =&\bigotimes_{j=1}^K{\left[ \left| \psi _{\mu _AS_{ij}} \right> \left| \psi _{\mu _BS_{ij}} \right> \right]}\otimes \bigotimes_{y=1}^K{\left[ \left| \psi _{\mu _AR_{xy}} \right> \left| \psi _{\mu _BR_{xy}} \right> \right]}
		\end{align}
		
\noindent\textbf{Step (3-3):}
		Based on the operation (7) of QZN, combine SQZV and obtain c-SQZV; combine RQZV and get c-RQZV.  c-SQZV and c-RQZV are defined as:
		\begin{align}
		\overrightarrow{\boldsymbol{CZ}_{\boldsymbol{S}_{\boldsymbol{i}}}^{\boldsymbol{Q}}}&=\left( \,\,CZ_{S_{i1}}^{Q},\,\,CZ_{S_{i2}}^{Q},\,\,...\,\,,\,\,CZ_{S_{ij}}^{Q},\,\,...\,\,,\,\,CZ_{S_{iK}}^{Q}\,\, \right) 
		\\
		\overrightarrow{\boldsymbol{CZ}_{\boldsymbol{R}_{\boldsymbol{x}}}^{\boldsymbol{Q}}}&=\left( \,\,CZ_{R_{x1}}^{Q},\,\,CZ_{R_{x2}}^{Q},\,\,...\,\,,\,\,CZ_{R_{xy}}^{Q},\,\,...\,\,,\,\,CZ_{R_{xK}}^{Q}\,\, \right) 
		\end{align}
		where $CZ_{S_{ij}}^{Q}$ and $ CZ_{R_{xy}}^{Q} $ are respectively the c-QZNs of samples and references:
		\begin{align}
		CZ_{S_{ij}}^{Q}&=\{\left. <\left| \varphi \right> ,\left| \psi _{\mu _{AB}S_{ij}} \right> >\,\, \right|\,\left| \varphi \right> \in \mathbb{U}\}
		\\
		CZ_{R_{xy}}^{Q}&=\{\left. <\left| \varphi \right> ,\left| \psi _{\mu _{AB}R_{xy}} \right> >\,\, \right|\,\left| \varphi \right> \in \mathbb{U}\}
		\end{align}
		in which $\left| \psi _{\mu _{AB}S_{ij}} \right>$ and $\left| \psi _{\mu _{AB}R_{xy}} \right>$ are the combined state for c-QZNs of samples and references:
		\begin{align}
		\left| \psi _{\mu _{AB}S_{ij}} \right> =&\sqrt{\mu _A\left( S_{ij} \right)}\sqrt{\mu _B\left( S_{ij} \right)}\left| 110 \right> 
		\notag\\
		&+\sqrt{\mu _A\left( S_{ij} \right)}\sqrt{1-\mu _B\left( S_{ij} \right)}\left| 101 \right> 
		\notag\\
		&+\sqrt{1-\mu _A\left( S_{ij} \right)}\sqrt{\mu _B\left( S_{ij} \right)}\left| 011 \right> 
		\notag\\
		&+\sqrt{1-\mu _A\left( S_{ij} \right)}\sqrt{1-\mu _B\left( S_{ij} \right)}\left| 001 \right> 
		\end{align}
		\begin{align}
		\left| \psi _{\mu _{AB}R_{xy}} \right> =&\sqrt{\mu _A\left( R_{xy} \right)}\sqrt{\mu _B\left( R_{xy} \right)}\left| 110 \right> 
		\notag\\
		&+\sqrt{\mu _A\left( R_{xy} \right)}\sqrt{1-\mu _B\left( R_{xy} \right)}\left| 101 \right> 
		\notag\\
		&+\sqrt{1-\mu _A\left( R_{xy} \right)}\sqrt{\mu _B\left( R_{xy} \right)}\left| 011 \right> 
		\notag\\
		&+\sqrt{1-\mu _A\left( R_{xy} \right)}\sqrt{1-\mu _B\left( R_{xy} \right)}\left| 001 \right> 
		\end{align}
		
		The implementations of this step in quantum circuit are as follows. Firstly, add $2K$ ancillary qubits $\ket{1}^{\otimes 2K}$ to $\ket{\phi_1}$. Secondly, apply $4K$ Pauli-X gate on $
		\left| \psi _{\mu _AS_{ij}} \right> $, $\left| \psi _{\mu _BS_{ij}} \right>$, $\left| \psi _{\mu _AR_{xy}} \right> $, and $\left| \psi _{\mu _BR_{xy}} \right> $. Finally, apply $2K$ CCNOT gate on $X\left( \left| \psi _{\mu _AS_{ij}} \right> \right) X\left( \left| \psi _{\mu _BS_{ij}} \right> \right) \left| 1 \right> $, $X\left( \left| \psi _{\mu _AR_{xy}} \right> \right) X\left( \left| \psi _{\mu _BR_{xy}} \right> \right) \left| 1 \right> $,  and obtain the quantum state $\ket{\phi_2}$ of c-SQZV and c-RQZV:
		\begin{align}
		\left| \phi _2 \right> =&\bigotimes_{j=1}^K{\left[ CCNOT\left( X\left( \left| \psi _{\mu _AS_{ij}} \right> \right) X\left( \left| \psi _{\mu _BS_{ij}} \right> \right) \left| 1 \right> \right) \right]}
		\notag\\
		&\otimes \bigotimes_{y=1}^K{\left[ CCNOT\left( X\left( \left| \psi _{\mu _AR_{xy}} \right> \right) X\left( \left| \psi _{\mu _BR_{xy}} \right> \right) \left| 1 \right> \right) \right]}
		\\
		=&\bigotimes_{j=1}^K{\left[ \left| \psi _{\mu _{AB}S_{ij}} \right> \right]}\otimes \bigotimes_{y=1}^K{\left[ \left| \psi _{\mu _{AB}R_{xy}} \right> \right]}
		\end{align}
		For convenience, let $\left| \psi _{\mu _{AB}S_i} \right>$ and $\left| \psi _{\mu _{AB}R_x} \right>$ denote that:
		\begin{align}
		\left| \psi _{\mu _{AB}S_i} \right> &=\bigotimes_{j=1}^K{\left| \psi _{\mu _{AB}S_{ij}} \right>}
		\\
		\left| \psi _{\mu _{AB}R_x} \right> &=\bigotimes_{y=1}^K{\left| \psi _{\mu _{AB}R_{xy}} \right>}
		\end{align}
		Then $\ket{\phi_2}$ can be written as $\ket{\phi_2}=\left| \psi _{\mu _{AB}S_i} \right> \left| \psi _{\mu _{AB}R_x} \right>$.
		
\noindent\textbf{Step (3-4):}
		Based on quantum fidelity \cite{nielsen2002quantum} and swap test \cite{buhrman2001quantum}, measure the similarity of c-SQZV and c-RQZV. Let the operation of controlled-SWAP be $CSWAP(\ket{x}\ket{\psi}\ket{\phi})$: $\ket{0}\ket{\psi}\ket{\phi} \rightarrow\ket{0}\ket{\psi}\ket{\phi} $ and $\ket{1}\ket{\psi}\ket{\phi} \rightarrow\ket{1}\ket{\phi}\ket{\psi} $, where the first one-dimensional qubit $\ket{x}$ is the control qubit, while $\ket{\psi}$ and $\ket{\phi}$ are the target qubits whose dimension can be larger than one.  
		In order to implementing this step in quantum circuit, firstly, add an ancillary qubit $\ket{0}$ to $\ket{\phi_2}$. Next, implement the circuit for quantum fidelity, and get the quantum state $\left| \phi _3 \right>$:
		\begin{align}
		\left| \phi _3 \right> =&\left( H\otimes I\otimes I \right) \otimes \left( CSWAP \right) 
		\notag\\
	    &\otimes \left( H\otimes I\otimes I \right) \otimes \left( \left| 0 \right> \left| \psi _{\mu _{AB}S_i} \right> \left| \psi _{\mu _{AB}R_x} \right> \right) 
		\notag\\
		=&\frac{1}{2}\left| 0 \right> \left( \left| \psi _{\mu _{AB}S_i} \right> \left| \psi _{\mu _{AB}R_x} \right> +\left| \psi _{\mu _{AB}R_x} \right> \left| \psi _{\mu _{AB}S_i} \right> \right) 
		\notag\\
		&+\frac{1}{2}\left| 1 \right> \left( \left| \psi _{\mu _{AB}S_i} \right> \left| \psi _{\mu _{AB}R_x} \right> -\left| \psi _{\mu _{AB}R_x} \right> \left| \psi _{\mu _{AB}S_i} \right> \right) 
		\end{align}
		
\noindent\textbf{Step (3-5):}
		Observe the ancillary qubit and obtain the probability $p(\ket{0})$:
		\begin{align}
		p\left( \left| 0 \right> \right) &=\left< \phi _3 \right|M_0^{\dag}M_0\left| \phi _3 \right> 
		\\
		&=\frac{1}{2}+\frac{1}{2}|\left< \psi _{\mu _{AB}S_i}|\psi _{\mu _{AB}R_x} \right> |^2
		\end{align}
		where $M_0=\left| 0 \right> \left< 0 \right|$ is the measurement operator, and $|\left< \psi _{\mu _{AB}S_i}|\psi _{\mu _{AB}R_x} \right> |^2$ is the quantum fidelity of $\ket{\psi _{\mu _{AB}S_i}}$ and $\ket{\psi _{\mu _{AB}R_x}}$.
		
\noindent\textbf{Step (3-6):}
		Calculate and output the quantum fidelity coefficient $F_{ix}$:
		\begin{align}
		F_{ix}=2 \times p\left( \left| 0 \right> \right) -1=|\left< \psi _{\mu _{AB}S_i}|\psi _{\mu _{AB}R_x} \right> |^2			
		\end{align}

\noindent\textbf{Step (4): Construct quantum fidelity matrix.}
	
	Construct quantum fidelity matrix (QFM) based on quantum fidelity coefficient $F_{ix}$:
	\begin{equation}
	QFM=\left[ \begin{smallmatrix}
	F_{11}&		F_{12}&		\cdots&		&		F_{1N}\\
	F_{21}&		\ddots&		&		&		F_{2N}\\
	\vdots&		&		F_{ix}&		&		\vdots\\
	&		&		&		\ddots&		\\
	F_{M1}&		F_{M2}&		\cdots&		&		F_{MN}\\
	\end{smallmatrix} \right] 
	\end{equation}
	
\noindent\textbf{Step (5): Find the index of the maximum value of $F_{ix}$ and make decision.}
	
	For each row  of QFM $(i = 1, 2, \, ... \, ,  M)$, find the index of the maximum value of quantum fidelity coefficient: 
	\begin{equation}
		x_i^{*} = \arg\underset{x}{\max}\left( F_{ix} \right)
	\end{equation}
	where $ x \in \{1,2, ... , N\}  $. After that, make multi-attributes decision: the $i$-th sample best matches the $x_i^{*}$-th reference. For example, if $F_{64}$ is the maximum value in the $6$-th row, then the $6$-th sample can be classified as the $4$-th reference.

\begin{figure}[t]
	\centering
	\includegraphics[scale=0.26]{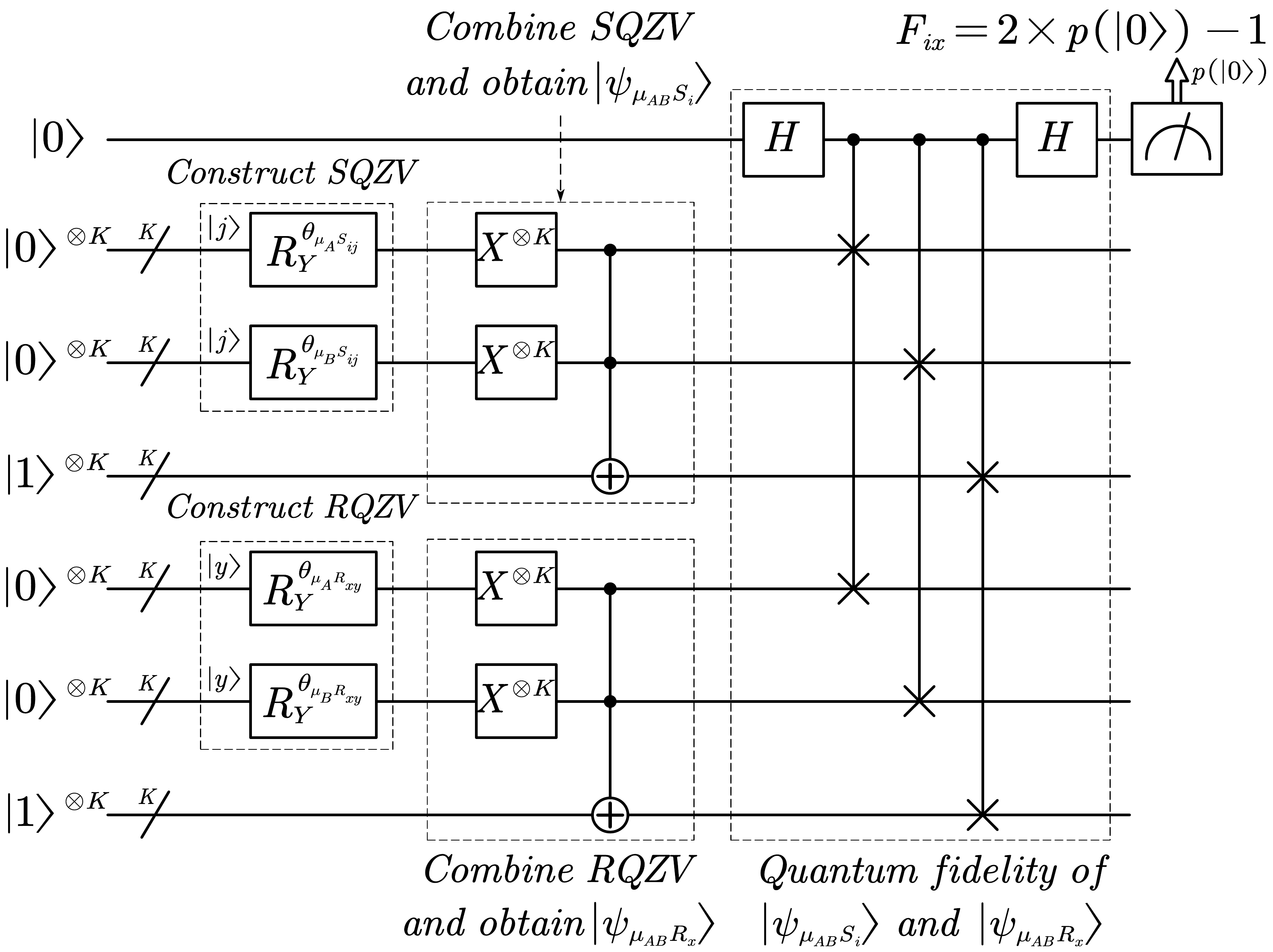}
	\caption{Quantum circuit for Step (3-1) to (3-6)}
	\label{Fig3}
\end{figure}

\begin{figure}[t]
	\centering
	\includegraphics[scale=0.43]{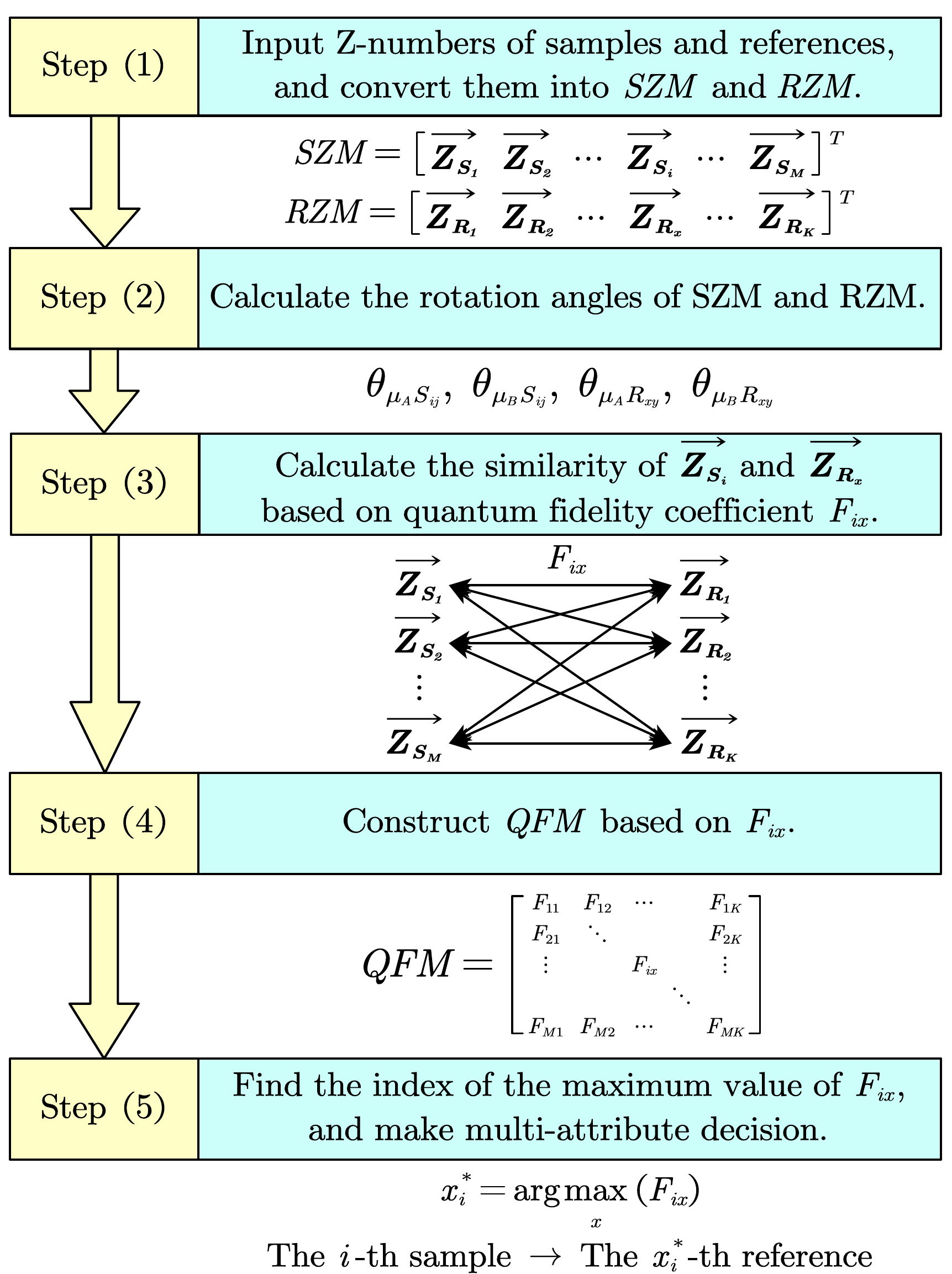}
	\caption{Procedure of the proposed algorithm}
	\label{Fig4}
\end{figure}

To illustrate, the implementation of the quantum circuit for Step (3-1) to (3-6) is illustrated in Fig. \ref{Fig3}. The overall procedure of the proposed algorithm is illustrated in Fig. \ref{Fig4}.

\subsection{Time complexity of the proposed algorithm}

In this subsection, firstly, the time complexity of the proposed algorithm is analyzed. Next, we compare the time complexity of the proposed algorithm with its classical counterpart.

In steps (1) and (2), SZM and RZM are prepared and processed in classical computer, and  the time cost is mainly taken for calculating the rotation angles of SZM and RZM.  SZM and RZM respectively  have  $MK$ and $NK$ elements, and each element  is a Z-number which has $2$ component. As a result, the  time cost of calculating the rotation angles is $2(M+N)K$.

\begin{figure}[h]
	\centering
	\includegraphics[scale=0.70]{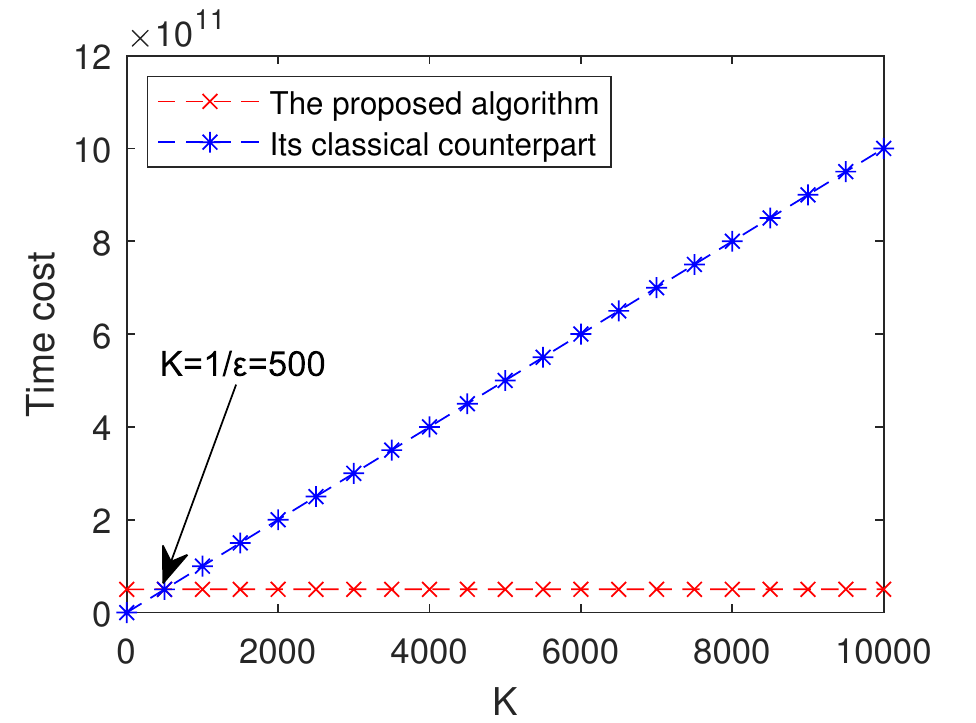}
	\caption{Time cost of the proposed method and that of its classical counterpart}
	\label{Fig5}
\end{figure}

Step (3) is implemented in quantum computer for $MN$ times until all $F_{ix}$ are calculated. 
As is shown in Fig. \ref{Fig3}, the Y-Rotation gates, Pauli-X gates, and CCNOT gates are respectively implemented in parallel, so that the time cost of them is $3$. Since each CSWAP gate works independently, these CSWAPs can be implemented in parallel, so that the time cost of them is also $1$. The time cost of the two Hadamard gate is $2$.  Hence, the time cost of the quantum circuit in Fig. \ref{Fig5} for one time is $6$. To achieve the desired error tolerance  $\varepsilon >0$, the quantum circuit in Fig. \ref{Fig5} should be independently implemented for $O(\frac{1}{\varepsilon})$ times \cite{buhrman2001quantum}. As a result,     the total time cost of step (3) is $O(\frac{1}{\varepsilon}MN)$. 

The main time cost of steps (4) and (5) is  taken to find the maximum value for each row of QFM, which is conducted in classical computer. QFM has $M$ rows, each row contains $N$ elements. The time cost for finding the maximum value among $N$ elements is $O(N)$ \cite{arora2009computational}. Therefore, the total time cost of steps (4) and (5) is $O(MN)$.

Based on above analysis,  the total time cost of the proposed algorithm is $O(\frac{1}{\varepsilon}MN+MK+NK)$.

Next, we will analyze the time complexity of the classical counterpart of the proposed algorithm. Since the classical counterpart is totally conducted in classical computer, the major difference between it and the proposed algorithm is in steps (3).

Step (3) measures the similarity of $K$-dimensional SZV and RZV based on quantum fidelity, so that its classical counterpart should also be a similarity-based or a distance-based algorithm. Several common similarity and distance measurements are Pearson correlation coefficient \cite{benesty2009pearson}, 
KL divergence \cite{kullback1951information}, and JS divergence \cite{lin1991divergence}. Given two $K$-dimensional vectors,  the time cost for these measurements of the two vectors  is at least $O(K)$ \cite{arora2009computational}. Because the step (3) is conducted for $MN$ times until the similarity of every SZV and RZV are calculated, the time cost for the classical counterpart of step (3) is $O(MNK)$. Hence, the total time cost of the classical counterpart of the proposed method is $O(MNK+MK+NK)$.

To illustrate the efficiency of the time complexity of the proposed method, the time cost of the proposed method and that of its classical counterpart are shown in Fig. \ref{Fig5} where the number of samples $M$ and the number of references $N$ are  $10000$, the error tolerance $\varepsilon$ is $0.002$, and the number of attributes $K$ increases from $1$ to $10000$. It can be seen that, when $K>\frac{1}{\varepsilon}$, with the  increase of $K$, the time cost of the proposed method is far less than that of its classical counterpart. 
Because $\frac{1}{\varepsilon}$ is irrelevant to the number of attributes $K$, the larger the $K$, the higher efficiency the quantum circuit is. Therefore,  blessed with the quantum parallelism, the proposed algorithm has the advantage of time complexity in big data scenario where $K$ is much larger than $\frac{1}{\varepsilon}$.


\section{Application in medical diagnosis}

In this section, the proposed algorithm will be applied in practical medical diagnosis problems. In addition, we will analyze and discuss about the proposed algorithm.

\subsection{Problem statement}

Assume  there are three patients: Alice, Bob, and Charlie denoted as $P_i$ where $i \in \{1, \, ... \, , 3\} $, and there are four symptoms: cough, temperature, headache,  and chest pain denoted as $SP_j$ where $j \in \{1, \, ... \, , 4\} $. Several diagnoses made by doctors are stomach problem, viral fever, malaria,  and typhoid denoted as $D_x$ where $x \in \{1, \, ... \, , 4\} $, and the associated  symptoms of the diagnoses are denoted as $SP_y$ where $y \in \{1, \, ... \, , 4\}$. These three patients can be seen as three samples, and these four diagnoses can be seen as four references. All of the samples and references have four symptoms which can be seen as four attributes.  Our goal is to match the three samples to the four references to make a medical diagnosis for every patient. 

Let $S$ denotes the UOD of all the patients (samples), and $S_{ij}\in S$ denotes the $i$-th patient and his or her $j$-th symptom (attribute), where $i \in \{1, \, ... \, , 3\} $ and $j \in \{1, \, ... \, , 4\}$.   Let $R$ denotes the UOD of all the diagnoses (references), and  $R_{xy}\in R$ denotes the $x$-th diagnosis and its $y$-th symptom (attribute), where $x \in \{1, \, ... \, , 4\} $ and $y \in \{1, \, ... \, , 4\}$.

The relationship between $P_i$ and $SP_j$ and their associated reliability are described by classical fuzzy membership function, which are shown in TABLE \ref{tab3} and TABLE \ref{tab4}. In addition,  the relationship between $D_x$ and $SP_y$ and their associated reliability are also described by classical fuzzy membership, which are shown in TABLE \ref{tab5} and TABLE \ref{tab6}.
\begin{table}[h]
	\centering
	\caption{Relationship between $P_i$ and $SP_j$}
	\setlength{\tabcolsep}{1mm}{
	\begin{tabular}{c|cccc}
		\hline
		& $SP_1$ & $SP_2$ & $SP_3$ & $SP_4$ \\
		\hline
		$P_1$ & $\left<S_{11},0.35\right>$ & $\left<S_{12},0.43\right>$ & $\left<S_{13},0.12\right>$ & $\left<S_{14},0.61\right>$ \\
		$P_2$ & $\left<S_{21},0.26\right>$ & $\left<S_{22},0.49\right>$ & $\left<S_{23},0.43\right>$ & $\left<S_{24},0.36\right>$ \\
		$P_3$ & $\left<S_{31},0.68\right>$ & $\left<S_{32},0.73\right>$ & $\left<S_{33},0.12\right>$ & $\left<S_{34},0.08\right>$ \\
		\hline
	\end{tabular}%
    }
	\label{tab3}%
\end{table}%
\begin{table}[h]
	\centering
	\caption{Reliability of  relationship between $P_i$ and $SP_j$}
	\setlength{\tabcolsep}{1mm}{
		\begin{tabular}{c|cccc}
			\hline
			& $SP_1$ & $SP_2$ & $SP_3$ & $SP_4$ \\
			\hline
			$P_1$ & $\left<S_{11},0.77\right>$ & $\left<S_{12},0.38\right>$ & $\left<S_{13},0.84\right>$ & $\left<S_{14},0.83\right>$ \\
			$P_2$ & $\left<S_{21},0.33\right>$ & $\left<S_{22},0.81\right>$ & $\left<S_{23},0.72\right>$ & $\left<S_{24},0.28\right>$ \\
			$P_3$ & $\left<S_{31},0.82\right>$ & $\left<S_{32},0.89\right>$ & $\left<S_{33},0.86\right>$ & $\left<S_{34},0.61\right>$ \\
			\hline
		\end{tabular}%
	}
	\label{tab4}%
\end{table}%
\begin{table}[h]
	\centering
	\caption{Relationship between $D_x$ and $SP_y$}
	\setlength{\tabcolsep}{1mm}{
		\begin{tabular}{c|cccc}
			\hline
			& $SP_1$ & $SP_2$ & $SP_3$ & $SP_4$ \\
			\hline
			$D_1$ & $\left<R_{11},0.41\right>$ & $\left<R_{12},0.43\right>$ & $\left<R_{13},0.37\right>$ & $\left<R_{14},0.12\right>$ \\
			$D_2$ & $\left<R_{21},0.84\right>$ & $\left<R_{22},0.86\right>$ & $\left<R_{23},0.21\right>$ & $\left<R_{24},0.15\right>$ \\
			$D_3$ & $\left<R_{31},0.25\right>$ & $\left<R_{32},0.32\right>$ & $\left<R_{33},0.69\right>$ & $\left<R_{34},0.38\right>$ \\
			$D_4$ & $\left<R_{41},0.18\right>$ & $\left<R_{42},0.24\right>$ & $\left<R_{43},0.14\right>$ & $\left<R_{44},0.79\right>$ \\
			\hline
		\end{tabular}%
	}
	\label{tab5}%
\end{table}%
\begin{table}[h]
	\centering
	\caption{Reliability of  relationship between $D_x$ and $SP_y$}
	\setlength{\tabcolsep}{1mm}{
		\begin{tabular}{c|cccc}
			\hline
			& $SP_1$ & $SP_2$ & $SP_3$ & $SP_4$ \\
			\hline
			$D_1$ & $\left<R_{11},0.83\right>$ & $\left<R_{12},0.87\right>$ & $\left<R_{13},0.81\right>$ & $\left<R_{14},0.82\right>$ \\
			$D_2$ & $\left<R_{21},0.95\right>$ & $\left<R_{22},0.92\right>$ & $\left<R_{23},0.87\right>$ & $\left<R_{24},0.85\right>$ \\
			$D_3$ & $\left<R_{31},0.91\right>$ & $\left<R_{32},0.96\right>$ & $\left<R_{33},0.89\right>$ & $\left<R_{34},0.92\right>$ \\
			$D_4$ & $\left<R_{41},0.81\right>$ & $\left<R_{42},0.87\right>$ & $\left<R_{43},0.84\right>$ & $\left<R_{44},0.85\right>$ \\
			\hline
		\end{tabular}%
	}
	\label{tab6}%
\end{table}%

\subsection{Application of the proposed algorithm}

In this subsection, the proposed QZN-based MADM algorithm is applied in medical diagnose. The calculating procedures are detailed as follows.

\noindent\textbf{Step (1):} Construct Z-numbers for samples and references, where the first component and the second component of Z-numbers for samples are respectively based on TABLE \ref{tab3} and TABLE \ref{tab4}, and that of Z-numbers for references respectively based on TABLE \ref{tab5} and TABLE \ref{tab6}. Then, convert these Z-numbers into SZM and RZM:
\begin{align}
SZM=\left[ \begin{smallmatrix}
\left< 0.35,0.77 \right>&		\left< 0.43,0.38 \right>&		\left< 0.12,0.84 \right>&		\left< 0.61,0.83 \right>\\
\left< 0.26,0.33 \right>&		\left< 0.49,0.81 \right>&		\left< 0.43,0.72 \right>&		\left< 0.36,0.28 \right>\\
\left< 0.68,0.82 \right>&		\left< 0.73,0.89 \right>&		\left< 0.12,0.86 \right>&		\left< 0.08,0.61 \right>\\
\end{smallmatrix} \right] 
\\
RZM=\left[ \begin{smallmatrix}
\left< 0.41,0.83 \right>&		\left< 0.43,0.87 \right>&		\left< 0.37,0.81 \right>&		\left< 0.12,0.82 \right>\\
\left< 0.84,0.95 \right>&		\left< 0.86,0.92 \right>&		\left< 0.21,0.87 \right>&		\left< 0.15,0.85 \right>\\
\left< 0.25,0.91 \right>&		\left< 0.32,0.96 \right>&		\left< 0.69,0.89 \right>&		\left< 0.38,0.92 \right>\\
\left< 0.18,0.81 \right>&		\left< 0.24,0.87 \right>&		\left< 0.14,0.84 \right>&		\left< 0.79,0.85 \right>\\
\end{smallmatrix} \right]  
\end{align}

\noindent\textbf{Step (2):} Calculate the corresponding rotation angles for every element of SZM and RZM, which are shown in TABLE \ref{tab7}.

\begin{table}[h]
	\centering
	\caption{Rotation angles of SZM and RZM}
	\setlength{\tabcolsep}{1mm}{
		\begin{tabular}{c|ccccc}
			\hline
			$\theta _{\mu _AS_{ij}}$ & $j=1$ & $j=2$ & $j=3$ & $j=4$ \\
			\hline
			$i=1$ & 107.458$^{\circ}$ & 98.048$^{\circ}$ & 139.464$^{\circ}$ & 77.291$^{\circ}$ \\
			$i=2$ & 118.685$^{\circ}$ & 91.146$^{\circ}$ & 98.048$^{\circ}$ & 106.260$^{\circ}$ \\
			$i=3$ & 68.900$^{\circ}$ & 62.613$^{\circ}$ & 139.464$^{\circ}$ & 147.140$^{\circ}$ \\
			\hline
			\hline
			$\theta _{\mu _BS_{ij}}$ & $j=1$ & $j=2$ & $j=3$ & $j=4$ \\
			\hline
			$i=1$ & 57.316$^{\circ}$ & 103.887$^{\circ}$ & 47.156$^{\circ}$ & 48.700$^{\circ}$ \\
			$i=2$ & 109.877$^{\circ}$ & 51.684$^{\circ}$ & 63.896$^{\circ}$ & 116.104$^{\circ}$ \\
			$i=3$ & 50.208$^{\circ}$ & 38.739$^{\circ}$ & 43.946$^{\circ}$ & 77.291$^{\circ}$ \\
			\hline
			\hline
			$\theta _{\mu _AR_{xy}}$ & $y=1$ & $y=2$ & $y=3$ & $y=4$ \\
			\hline
			$x=1$ & 100.370$^{\circ}$ & 98.048$^{\circ}$ & 105.070$^{\circ}$ & 139.464$^{\circ}$ \\
			$x=2$ & 47.156$^{\circ}$ & 43.946$^{\circ}$ & 125.451$^{\circ}$ & 134.427$^{\circ}$ \\
			$x=3$ & 120.000$^{\circ}$ & 111.100$^{\circ}$ & 67.666$^{\circ}$ & 103.887$^{\circ}$ \\
			$x=4$ & 129.792$^{\circ}$ & 121.332$^{\circ}$ & 136.054$^{\circ}$ & 54.549$^{\circ}$ \\
			\hline
			\hline
			$\theta _{\mu _BR_{xy}}$ & $y=1$ & $y=2$ & $y=3$ & $y=4$ \\
			\hline
			$x=1$ & 48.700$^{\circ}$ & 42.269$^{\circ}$ & 51.684$^{\circ}$ & 50.208$^{\circ}$ \\
			$x=2$ & 25.842$^{\circ}$ & 32.860$^{\circ}$ & 42.269$^{\circ}$ & 45.573$^{\circ}$ \\
			$x=3$ & 34.915$^{\circ}$ & 23.074$^{\circ}$ & 38.739$^{\circ}$ & 32.860$^{\circ}$ \\
			$x=4$ & 51.684$^{\circ}$ & 42.269$^{\circ}$ & 47.156$^{\circ}$ & 45.573$^{\circ}$ \\
			\hline
		\end{tabular}%
	}
	\label{tab7}%
\end{table}%

\noindent\textbf{Step (3):}  For one SZV $\overrightarrow{\boldsymbol{Z}_{\boldsymbol{S}_{\boldsymbol{i}}}}$  and one RZV $\overrightarrow{\boldsymbol{Z}_{\boldsymbol{R}_{\boldsymbol{x}}}}$, calculate their similarity based on quantum fidelity coefficient $F_{ix}$, until all $F_{ix}$ are obtained, where $ i \in \{1, ... , 3\}  $ and $ x \in \{1, ... , 4\}  $. Each $F_{ix}$ is calculated by the quantum circuit shown in Fig. \ref{Fig3},  
which is simulated in Qiskit and implemented in the quantum computer of IBM. 

Take $\overrightarrow{\boldsymbol{Z}_{\boldsymbol{S}_{\boldsymbol{1}}}}$ and $\overrightarrow{\boldsymbol{Z}_{\boldsymbol{R}_{\boldsymbol{2}}}}$ as an example.  Their corresponding rotation angles are $\theta _{\mu _AS_{1j}}$, $\theta _{\mu _BS_{1j}}$ $(j = 1, ... , 4 )  $ and $\theta _{\mu _AR_{2y}}$, $\theta _{\mu _BR_{2y}}$ $(y = 1, ... , 4 )  $, which are  shown in TABLE \ref{tab7}. Then, assign these rotation angles to their associated Y-Rotation gates, and create the quantum circuit for calculating the quantum fidelity coefficient $F_{12}$ of $\overrightarrow{\boldsymbol{Z}_{\boldsymbol{S}_{\boldsymbol{1}}}}$ and $\overrightarrow{\boldsymbol{Z}_{\boldsymbol{R}_{\boldsymbol{2}}}}$.  
Next, simulate the quantum circuit and get the probability: $p(\ket{0}) = 0.6316$.
Finally, the quantum fidelity coefficient of $\overrightarrow{\boldsymbol{Z}_{\boldsymbol{S}_{\boldsymbol{1}}}}$ and $\overrightarrow{\boldsymbol{Z}_{\boldsymbol{R}_{\boldsymbol{2}}}}$ can be calculated: $F_{12}=2 \times p\left( \left| 0 \right> \right) -1 = 0.2632$.

\noindent\textbf{Step (4):} Construct the quantum fidelity matrix (QFM) based on $F_{ix}$ in step (3), which is shown as follows:
\begin{equation}
QFM=\left[ \begin{smallmatrix}
0.5046 & 0.2632 & 0.3366 & \boldsymbol{0.6434} \\
\boldsymbol{0.4382} & 0.1768 & 0.2708 & 0.3198 \\
0.7194 & \boldsymbol{0.8146} & 0.3282 & 0.2520 \\
\end{smallmatrix} \right]
\label{QFM} 
\end{equation}

\noindent\textbf{Step (5):}  For each row of QFM, find  the maximum value of $F_{ix}$, which are  highlighted in bold in Eq. \ref{QFM}, and then make medical diagnosis based on the index of the maximum value. The medical diagnosis generated by the proposed algorithm is shown in the first row of TABLE \ref{tab8}.
\begin{table}[h]
	\centering
	\caption{Medical diagnoses made by three algorithms}
	\setlength{\tabcolsep}{0.8mm}{
		\begin{tabular}{l|lll}
			\hline
			& Alice & Bob   & Charlie \\
			\hline
			Proposed algorithm & Typhoid & Stomach problem & Viral fever \\
			ZN-based algorithm & Typhoid & Stomach problem & Viral fever \\
			QFS-based algorithm & Typhoid & Malaria & Viral fever \\
			\hline
		\end{tabular}%
	}
	\label{tab8}%
\end{table}%

\subsection{Analysis and discussion}

In this subsection, we analyze and discuss about the proposed QZN-based algorithm compared with ZN-based algorithm and QFS-based algorithm.

\noindent\emph{1) Compared with ZN-based algorithm}

For comparing QZN  with Z-number (ZN), the ZN-based algorithm is as follows, which is a classical counterpart of the proposed QZN-based algorithm.   

\noindent\textbf{Step (1):} Input Z-numbers of samples of references, and construct SZM and RZM. 

\noindent\textbf{Step (2):} Construct combined-SZM and combined-RZM:
\begin{align}
CSZM&=\left[ \begin{smallmatrix}
\overrightarrow{\boldsymbol{CZ}_{\boldsymbol{S}_{\boldsymbol{1}}}}&		\overrightarrow{\boldsymbol{CZ}_{\boldsymbol{S}_{\boldsymbol{2}}}}&		\cdots&		\overrightarrow{\boldsymbol{CZ}_{\boldsymbol{S}_{\boldsymbol{i}}}}&		\cdots&		\overrightarrow{\boldsymbol{CZ}_{\boldsymbol{S}_{\boldsymbol{M}}}}\\
\end{smallmatrix} \right] ^T
\\
CRZM&=\left[ \begin{smallmatrix}
\overrightarrow{\boldsymbol{CZ}_{\boldsymbol{R}_{\boldsymbol{1}}}}&		\overrightarrow{\boldsymbol{CZ}_{\boldsymbol{R}_{\boldsymbol{2}}}}&		\cdots&		\overrightarrow{\boldsymbol{CZ}_{\boldsymbol{R}_{\boldsymbol{x}}}}&		\cdots&		\overrightarrow{\boldsymbol{CZ}_{\boldsymbol{R}_{\boldsymbol{N}}}}\\
\end{smallmatrix} \right] ^T
\end{align}
in which
\begin{align}
\overrightarrow{\boldsymbol{CZ}_{\boldsymbol{S}_{\boldsymbol{i}}}}&=\left( \begin{smallmatrix}  CZ_{S_{i1}},&CZ_{S_{i2}},&...&,CZ_{S_{ij}},&...&,CZ_{S_{iK}}\\  \end{smallmatrix} \right) \\
\overrightarrow{\boldsymbol{CZ}_{\boldsymbol{R}_{\boldsymbol{x}}}}&=\left( \begin{smallmatrix}  CZ_{R_{x1}},&CZ_{R_{x2}},&...&,CZ_{R_{xy}},&...&,CZ_{R_{xK}}\\  \end{smallmatrix} \right)
\end{align} 
are called the combined-SZV and combined-RZV, where $CZ_{S_{ij}}=$$\mu _A\left( S_{ij} \right)  \mu _B\left( S_{ij} \right)  $ and $CZ_{R_{xy}}=$$ \mu _A\left( R_{xy} \right)  \mu _B\left( R_{xy} \right) $.

\noindent\textbf{Step (3):} Given one combined-SZV $\overrightarrow{\boldsymbol{CZ}_{\boldsymbol{S}_{\boldsymbol{i}}}}$ and one combined-RZV $\overrightarrow{\boldsymbol{CZ}_{\boldsymbol{R}_{\boldsymbol{x}}}}$, measure their similarity between  based on Pearson correlation coefficient \cite{benesty2009pearson}:
\begin{equation}
r_{ix}=\frac{cov(\overrightarrow{\boldsymbol{CZ}_{\boldsymbol{S}_{\boldsymbol{i}}}}, \overrightarrow{\boldsymbol{CZ}_{\boldsymbol{R}_{\boldsymbol{x}}}})}{\sigma_{\overrightarrow{\boldsymbol{CZ}_{\boldsymbol{S}_{\boldsymbol{i}}}}}\sigma_{\overrightarrow{\boldsymbol{CZ}_{\boldsymbol{R}_{\boldsymbol{x}}}}}}
\end{equation}
where $cov$$(\overrightarrow{\boldsymbol{CZ}_{\boldsymbol{S}_{\boldsymbol{i}}}}, \overrightarrow{\boldsymbol{CZ}_{\boldsymbol{R}_{\boldsymbol{x}}}})$ is covariance, and  $\sigma_{\overrightarrow{\boldsymbol{CZ}_{\boldsymbol{S}_{\boldsymbol{i}}}}}$   $\sigma_{\overrightarrow{\boldsymbol{CZ}_{\boldsymbol{R}_{\boldsymbol{x}}}}}$ are respectively the standard deviation of $\overrightarrow{\boldsymbol{CZ}_{\boldsymbol{S}_{\boldsymbol{i}}}}$ and  $\overrightarrow{\boldsymbol{CZ}_{\boldsymbol{R}_{\boldsymbol{x}}}}$.

\noindent\textbf{Step (4):} Construct Pearson correlation coefficient matrix (PM) based on $r_{ix}$:
\begin{equation}
PM=\left[ \begin{smallmatrix}
r_{11}&		\cdots&		r_{1N}\\
\vdots&		\ddots&		\vdots\\
r_{M1}&		\cdots&		r_{MN}\\
\end{smallmatrix} \right]
\end{equation}

\noindent\textbf{Step (5):} For each row of PM, find the index of the maximum value of $r_{ix}$ and make decision.

Then, this ZN-based algorithm is applied in medical diagnosis, and the PM is calculated as:
\begin{equation}
PM=\left[ \begin{smallmatrix}
-0.8524 & -0.3361 & -0.4371 & \boldsymbol{0.9198} \\
\boldsymbol{0.5588} & 0.1624 & 0.4158 & -0.4643 \\
0.7928 & \boldsymbol{0.9915} & -0.6616 & -0.5285 \\
\end{smallmatrix} \right] 
\end{equation}
where the maximum values of each row are in bold. Next, the medical diagnosis generated by the ZN-based algorithm is shown in the second row of TABLE \ref{tab8}. 

The diagnosis for each patient made by the ZN-based algorithm  is the same as  that made by the proposed algorithm, which shows that both of these two algorithms can effectively handle fuzziness and make correct medical diagnosis for different patients. However, as is discussed in Subsection $B$ of Section \uppercase\expandafter{\romannumeral5}, since the ZN-based algorithm is based on Pearson correlation coefficient, its total time cost is much slower than that of the proposed algorithm in big data scenario. In addition, the ZN-based algorithm is a classical algorithm, which cannot be applied in quantum information processing.

By comparison, assisted by quantum computation, the proposed QZN-based algorithm can efficiently make medical diagnosis. Moreover, it can be extended and applied in quantum information processing under other scenario.

\noindent\emph{2) Compared with QFS-based algorithm}

To compare QZN with quantum fuzzy set (QFS), QZN in the proposed algorithm is replaced by QFS, while the other procedures of the proposed algorithm remain the same. This algorithm is called the QFS-based algorithm. Then, we compare the proposed QZN-based algorithm with the QFS-based algorithm.   The quantum fidelity matrix (QFM) generated by the QFS-based algorithm is that
\begin{equation}
QFM=\left[ \begin{smallmatrix}
0.6554 & 0.4514 & 0.6030 & \boldsymbol{0.8792} \\
0.8830 & 0.4908 & \boldsymbol{0.9044} & 0.6572 \\
0.7474 & \boldsymbol{0.9112} & 0.3896 & 0.2790 \\
\end{smallmatrix} \right]
\end{equation}
where the maximum values of each row are in bold. Then, the medical diagnosis made by the QFS-based algorithm is shown in the third row of TABLE \ref{tab8}. 

It can be seen that the diagnosis of Bob generated by the QFS-based algorithm is different from that of the proposed algorithm. The reason is that the QFS-based algorithm does not take reliability of quantum fuzziness into consideration, so that it makes wrong diagnosis. To be specific, the QFS-based algorithm cannot use the reliability information in TABLE \ref{tab4} and TABLE \ref{tab6}. Its omits that the reliability for the first and the fourth symptoms of  Bob are 0.33 and 0.28, which means that the fuzzy relationships for the first and the fourth symptoms of Bob should not be fully trusted.

By contrast, with the help of quantum sets $A^Q$ and $B^Q$, QZN can not only use $A^Q$ to represent the quantum fuzzy restriction of the elements in QUOD, but also express  the reliability of $A^Q$ based on $B^Q$, so that the proposed algorithm can make diagnoses correctly and efficiently.

Based on above discussion, compared with ZN-based algorithm and QFS-based algorithm,  the advantages of the proposed algorithm are summarized in TABLE \ref{tab9}.   

\begin{table}[h]
	\centering
	\caption{Comparison of three algorithms}
	\setlength{\tabcolsep}{1mm}{
		\begin{tabular}{l|ccc}
			\hline
			&   &  Efficiency  &   Quantum  \\
			&  & of time  &  information \\
			&Correctness & complexity &  processing \\
			\hline
			Proposed algorithm &  \checkmark &  \checkmark &  \checkmark \\
			ZN-based algorithm &  \checkmark &  $\times$ & $\times$  \\
			QFS-based algorithm & $\times$  &  \checkmark &  \checkmark \\
			\hline
		\end{tabular}%
	}
	\label{tab9}%
\end{table}%

\section{Conclusion}

Z-number proposed by Zadeh is an efficient tool for modeling  uncertainty in fuzzy environment. However, Z-numbers are unable to deal with uncertain information in quantum field. Therefore,  in order to equipping Z-numbers with the ability of processing quantum information, this paper generalizes Z-number into its quantum counterpart and proposes quantum Z-numbers (QZNs).

The main contributions of this paper are summarized as follows. (1) Quantum Z-numbers (QZNs) are proposed, which are the quantum extension of classical Z-numbers. A QZN consists of two quantum fuzzy sets,  taking both quantum fuzzy restriction  and its reliability into consideration.   (2) We present several basic quantum fuzzy  operations  of QZNs and their associated quantum circuits, which are expounded by numerical examples. (3)  A novel QZN-based quantum MADM algorithm is proposed. The analysis shows that the proposed algorithm has the advantage of time complexity  in big data scenario.  (4) The proposed algorithm is applied in medical diagnosis, which shows that the proposed algorithm can not only process fuzzy information efficiently but also make diagnoses correctly with low time complexity. 

In the future research, we will focus on designing other quantum algorithms of QZNs, such as quantum ranking of QZNs. Besides, applying QZN and its quantum algorithms into more practical fields is also worth exploring.


%

%

\section*{Acknowledgment}

The work is partially supported by National Natural Science Foundation of China (Grant No. 61973332), Invitational Fellowships for Research in Japan (Short-term).

\ifCLASSOPTIONcaptionsoff
  \newpage
\fi



\bibliographystyle{IEEEtran}
\bibliography{IEEEabrv.bib}
\end{document}